\documentclass[prx,aps,showpacs,twocolumn,preprintnumbers,
amsmath,amssymb,superscriptaddress,longbibliography]{revtex4-2}
\usepackage[english]{babel}
\usepackage{amsmath,amssymb,amsfonts}
\usepackage{graphicx}
\usepackage[colorlinks=True,linkcolor=red,citecolor=blue,urlcolor=blue]{hyperref}
\usepackage{bookmark}
\usepackage{xcolor}
\usepackage{braket}
\usepackage{nicefrac}
\usepackage{bm}
\usepackage{bbm}
\usepackage{braket}
\usepackage{slashed}
\usepackage[normalem]{ulem}
\usepackage{array}
\usepackage{makecell}
\newcolumntype{C}{>{$}c<{$}}

\def\id{\mathbbm{1}}

\newcommand{\tr}{\mathop{\mathrm{tr}}}

\renewcommand{\Re}{\mathop{\mathrm{Re}}}
\renewcommand{\Im}{\mathop{\mathrm{Im}}}
\newcommand{\R}{\mathbb{R}}
\renewcommand{\i}{\mathop{\mathrm{i}}}

\newcommand{\bolds}[1]{\boldsymbol #1}

\allowdisplaybreaks

  \DeclareUnicodeCharacter{2212}{-}

\begin{document}

\title{Realizing exceptional points of any order in the presence of symmetry
}
\author{Sharareh Sayyad}
\email{sharareh.sayyad@mpl.mpg.de}
\author{Flore K. Kunst}
\email{flore.kunst@mpl.mpg.de}
\affiliation{Max Planck Institute for the Science of Light, Staudtstra\ss e 2, 91058 Erlangen, Germany}
\date{\today}

\begin{abstract}
Exceptional points~(EPs) appear as degeneracies in the spectrum of non-Hermitian matrices at which the eigenvectors coalesce. In general, an EP of order $n$ may find room to emerge if $2(n-1)$ real constraints are imposed. Our results show that these constraints can be expressed in terms of the determinant and traces of the non-Hermitian matrix. Our findings further reveal that the total number of constraints may reduce in the presence of unitary and antiunitary symmetries. Additionally, we draw generic conclusions for the low-energy dispersion of the EPs. Based on our calculations, we show that in odd dimensions the presence of sublattice or pseudo-chiral symmetry enforces $n$th order EPs to disperse with the $(n-1)$th root.
For two-, three- and four-band systems, we explicitly present the constraints needed for the occurrence of EPs in terms of system parameters and classify EPs based on their low-energy dispersion relations. 
\end{abstract}

\maketitle

\section{Introduction}

The appearance of symmetry-protected degeneracies in the energy dispersion of various Hermitian topological systems has attracted much attention in the past decades~\cite{Wan2011, Burkov2011, Wang2012, Chiu2014, Senthil2015, Liang2016, Chiu2016, Bansil2016}. These Hermitian topological systems, aside from their space-group symmetry, are classified using ten symmetry classes~\cite{Altland1997} identified based on three discrete symmetries, namely time-reversal symmetry, particle-hole~(or charge conjugation) symmetry, and chiral~(or sublattice) symmetry~\cite{Chiu2016}. Topological semimetals~\cite{Xu2011, Young2012, Armitage2018} and multifold fermions~\cite{Bradlyn2016, Flicker2018, Tian2021} are excellent representatives of such systems in which two- or multi-band crossings can be observed in the energy spectra. In the absence of symmetry, these band touchings 
are generally unstable in lower-dimensional models due to the hybridization of the bands resulting in the gaping out of degeneracies. However, this band repulsion mechanism is absent in topological systems in which crystalline symmetries and/or discrete symmetries, e.g., time-reversal symmetry, may protect band touching points~\cite{Wu2020}.

It has further been shown that the commonly observed linear energy dispersion close to nontrivial degeneracies might be forbidden due to certain symmetry constraints present in some systems~\cite{Fang2012}. As a result, higher-order band dispersions, such as cubic or quadratic, may find room to arise close to band touching manifolds~\cite{ Yu2018cub, Yu2019c, Zhang2021}. These distinct characters of energy spectra are considered as an additional tool to classify various nontrivial degeneracies in Hermitian systems~\cite{Wu2020}.

The recent surge of theoretical and experimental interests in the field of non-Hermitian systems has advanced our understanding of the intrinsic properties of systems with no Hermitian counterparts. Some of these exotic properties are i) the piling up of bulk states on the boundaries known as the non-Hermitian skin effect~\cite{Yao2018skin}, which goes hand in hand with a violation of the conventional (Hermitian) bulk-boundary correspondence~\cite{Lee2016, Kunst2018b, Kunst2019}, ii) the emergence of exceptional points~(EPs)~\cite{Kawabata2019, Bergholtz2021} as defective degeneracies at which the geometric multiplicity is smaller than the algebraic multiplicity, and iii) the observation of different non-Hermitian topological systems due to the closure of non-Hermitian (line or point) gaps~\cite{Yao2018chern, Yin2018, Sayyad2021}.

The emergence of these unique properties of non-Hermitian systems is linked to the extended 38 symmetry classes~\cite{Kawabata2019}, which are the non-Hermitian counterparts of the tenfold Altland–Zirnbauer classification~\cite{Altland1997, Chiu2016} in Hermitian systems. As Hermiticity is not respected in non-Hermitian systems, particle-hole symmetry~($\rm PHS$ and ${\rm PHS}^{\dagger}$) and time-reversal symmetry~($\rm TRS$ and ${\rm TRS}^{\dagger}$) acquire two different flavors, and chiral symmetry~($\rm CS$) is discerned from sublattice symmetry~($\rm SLS$). These six symmetries combined with psuedo-Hermiticity~($\rm psH$)~\cite{Mostafazadeh2002} give rise to 38 symmetry classes as defined in Ref.~\cite{Kawabata2019}. Aside from these seven symmetries, pseudo-chiral symmetry~($\rm psCS$)~\cite{Rivero2020}, inversion~($\cal I$) symmetry~\cite{Yoshida2021}, parity~($\cal P$) symmetry and its combination with time-reversal~($\cal PT$)~\cite{Bender1998} and particle-hole~($\cal CP$) symmetries~\cite{Delplace2021} have been considered in exploring various properties of non-Hermitian systems. We summarize these symmetries in Table~\ref{tab:symm}. We note that these symmetries can also be written in terms of the classification of random matrices~\cite{Lee2019} introduced by Bernard and LeClair~\cite{Bernard2002}, see also Appendix~\ref{app:symmBL} for details. 

Among the unique properties of non-Hermitian systems, acquiring a deeper understanding regarding exceptional points has been the focus of numerous recent theoretical~\cite{ Yuce2018, Okugawa2019, Budich2019,  Stalhammar2021, Bergholtz2021, Crippa2021, Yang2021, Delplace2021} and experimental~\cite{Hao2018, Miri2019, Dembowski2001, Sakhdari2021, Ding2021} studies because of their putative applications, for instance, in sensing devices~\cite{Kazemi2020, Wiersig2020} and unidirectional lasing~\cite{Longhi2017, Huang2017}. While the major focus of these works has been mainly on exceptional points of order two, i.e., exceptional points at which two eigenvalues coincide and simultaneously associated eigenvectors coalesce onto one, a recent shift has been made towards studying the properties of EPs with higher orders~\cite{Budich2019, Okugawa2019, Mandal2021, Crippa2021, Stalhammar2021, Delplace2021}.

These investigations, which usually explore case studies, mainly address a number of questions as follows: i)~How many constraints need to be satisfied to find an $n$th order EP, dubbed as EP$n$?
It has been argued that $2(n-1)$ real constraints should be imposed to detect EP$n$s in systems with no symmetry~\cite{Bergholtz2021}. Even though a description for these constraints is discussed in Ref.~\onlinecite{Delplace2021}, a generic recipe to generate and understand these constraints in the presence of any symmetry is absent in the literature. Nevertheless, it has been suggested that relating each of these constraints to a momentum coordinate implies that merely EP2s can be realized in three spatial dimensions~\cite{Berry2004, Carlstrom2018}.
ii)~What role is played by symmetries in the appearance of EP$n$s? Recent researches reveal that including symmetries may reduce the number of constraints to realize EPs. As a result, various case studies reported the occurrence of Ep2s~\cite{Budich2019, Okugawa2019}, Ep3s~\cite{Mandal2021}, and Ep4s~\cite{Crippa2021} in one, two, and three spatial dimensions, respectively. More extended studies have also explored the link between observing EP$n$s and the presence of either $\mathcal{PT}$~\cite{Stalhammar2021} or antiunitary~\cite{Delplace2021} symmetries.
iii)~Is it possible to distinguish EP$n$s based on the low-energy dispersions close to them?
Similar to Hermitian systems at which linear, cubic, and quadratic dispersions were reported close to nontrivial degeneracies, $n$th-root dispersion in the vicinity of EP$n$s were numerously identified~\cite{Budich2019}. A recent study reports the square-root behavior of band spectra close to EP$3$s in the presence of SLS~\cite{Mandal2021}, where this possibility has also been studied in Ref.~\onlinecite{Demange2011} without reference to symmetry.

In this work, we revisit these questions using a generic mathematical formulation to explore the appearance of EP$n$s in Hamiltonians represented by $n$-dimensional matrices. Based on our formalism, we are able to count the number of constraints in the presence of any symmetry and evaluate each constraint based on the traces and the determinant of the Hamiltonian of our interests. In particular, we find that one needs to satisfy $\tr[\mathcal{H}^k] = 0$ with $k = 2, \ldots, n-1$ and $\det[\mathcal{H}] = 0$ to find an EP with order $n$ arriving at a total of $2(n-1)$ constraints in agreement with the literature.
Imposing symmetry considerations, we show that in the presence of CS, psCS, SLS, psH, $\cal PT$, and $\cal CP$ symmetries, some traces or the determinant of $n$-band systems generally disappear. Moreover, when psH, $\mathcal{PT}$ or $\mathcal{CP}$ symmetry is present, we find that the number of constraints is reduced to half, i.e., $n-1$ constraints.
When we instead consider psCS or SLS, we recover $n$ constraints for $n \in \textrm{even}$ and $n-1$ constraints when $n \in \textrm{odd}$. CS is only defined in even dimensions, in which case we find $n-1$ constraints. We summarize these results in Table~\ref{tab:symm_epn}.

We, furthermore, identify conditions to characterize various EP$n$s based on their low-energy dispersions.
To do so, we introduce an alternative approach based on the Frobenius companion matrix of the characteristic polynomial, which can be interpreted as representing a perturbation close to an EP$n$. With this matrix in mind, we rederive the above statement pertaining to the $2(n-1)$ constraints as well as explicitly calculate the low-energy band dispersions around an EP$n$.
Despite the common assumption that EP$n$s disperse with the $n$th root, we find that in the presence of SLS or psCS with $n \in \textrm{odd}$, the leading order term of the dispersion around an EP$n$ generically scales with the $(n-1)$th root.

We emphasize that our formulation is not limited to any specific spatial dimension. For completeness purposes, we calculate explicit forms for the nonzero constraints for all twelve symmetries listed in Table~\ref{tab:symm} for two-~, three-~, and four-band systems and present their nonzero parameters.

The outline of this paper is as follows. In Sec.~\ref{sec:nEps}, we present our generic mathematical formulation to describe EP$n$s. We further draw generic symmetry-based arguments on the behavior of EP$n$s when a specific symmetry is respected. Using the generic decomposition of two-band systems in terms of Pauli matrices, we discuss the properties of EP$2$s, explicit forms of constraints, and collections of nonzero parameters in the presence of each twelve symmetries in Sec.~\ref{sec:2lvl}. In Sections \ref{sec:3lvl} and \ref{sec:4lvl} we pursue similar lines of thought for EP$3$s and EP$4$s, respectively. Using the Gell-Mann matrices and their generalization, we rewrite three- and four-band Hamiltonians and identify their nonzero components when a symmetry constraint is enforced. We also discuss various possibilities to observe different energy dispersions close to EP$3$s and EP$4$s in Sections~\ref{sec:3lvl} and \ref{sec:4lvl}, respectively. We conclude our paper in Sec.~\ref{sec:conclusion}.

\begin{table*}
     \centering
    \caption{Summarized symmetries and their associated energy constraints}
     \begin{tabular}{l|ll|ll}
        \hline \hline 
        Symmetry  &\qquad & Symmetry constraint  &\qquad  &Energy constraint  \\
        \hline 
        \hline
        Particle-hole symmetry I~(PHS)   
        &\qquad  & ${\cal H}(-\bolds{k}) =- {\cal C}_{-} {\cal H}^{T}(\bolds{k}) {\cal C}_{-}^{\dagger} $     
       &\qquad  &   $\{\epsilon(\bolds{k})\} = \{- \epsilon(-\bolds{k}) \}$  
        \\
        Particle-hole symmetry II~(PHS$^{\dagger}$)    
        & \qquad & ${\cal H}(-\bolds{k}) = -{\cal T}_{-} {\cal H}^{*}(\bolds{k}) {\cal  T}_{-}^{\dagger} $       
        & \qquad &   $\{\epsilon(\bolds{k})\} = \{- \epsilon^{*}(-\bolds{k})\}$  
        \\
        Time-reversal symmetry I~(TRS)    
        & \qquad & ${\cal H}(-\bolds{k}) = {\cal T}_{+} {\cal H}^{*}(\bolds{k}) {\cal T}_{+}^{\dagger} $        
        & \qquad &  $\{\epsilon(\bolds{k})\} = \{ \epsilon^{*}(-\bolds{k})\}$   \\
        Time-reversal symmetry II~(TRS$^{\dagger}$)    
        & \qquad & ${\cal H}(-\bolds{k}) = {\cal C}_{+} {\cal H}^{T}(\bolds{k}) {\cal C}_{+}^{\dagger} $       
        & \qquad &  $\{\epsilon(\bolds{k}) \} =   \{ \epsilon(-\bolds{k})\}$
 \\
        Chiral symmetry~(CS)  
        & \qquad & $  {\cal H}(\bolds{k}) = - \Gamma {\cal H}^{\dagger}(\bolds{k}) \Gamma^{-1} $    
        & \qquad &  $ \{\epsilon(\bolds{k})\} = \{- \epsilon^{*}(\bolds{k})\}$
\\
        Pseudo-chiral symmetry~(${\rm ps CS}$) 
        & \qquad & $  {\cal H}^{T}(\bolds{k}) = - \Lambda {\cal H}(\bolds{k}) \Lambda^{-1} $    
        & \qquad &  $ \{\epsilon(\bolds{k}) \}= \{- \epsilon(\bolds{k})\}$
\\
                 Sublattice-symmetry~(SLS)  
        & \qquad & ${\cal H}(\bolds{k}) = - {\cal S} {\cal H}(\bolds{k}) {\cal S}^{-1}  $  
        & \qquad &  $\{ \epsilon(\bolds{k}) \}= \{- \epsilon(\bolds{k})\}$ 
 \\
                Pseudo-Hermiticity~($\rm psH$) 
        & \qquad & ${\cal H}(\bolds{k}) = \varsigma {\cal H}^{\dagger}(\bolds{k}) \varsigma^{-1} $  
        & \qquad &  $\{ \epsilon(\bolds{k}) \} =\{ \epsilon^{*}(\bolds{k})\}$
\\
               Inversion symmetry~($\cal I$)
        & \qquad & ${\cal H}^{\dagger}(-\bolds{k}) = {\cal I} {\cal H}(\bolds{k}) {\cal I}^{-1}  $ 
        & \qquad &  $\{\epsilon(\bolds{k} ) \} =\{ \epsilon^{*}(-\bolds{k})\}$
 \\
         Parity~($\cal P$) symmetry
        & \qquad & ${\cal H}(-\bolds{k}) ={\cal P} {\cal H}(\bolds{k}) {\cal P}^{-1}  $    
        & \qquad &  $\{\epsilon(\bolds{k}) \} =\{ \epsilon(-\bolds{k})\}$
 \\
         Parity-time~($\cal PT$) symmetry
        & \qquad & ${\cal H}(\bolds{k}) =({\cal P}{\cal T}_{+}) {\cal H}^{*}(\bolds{k}) ({\cal P}{\cal T}_{+})^{-1} $   
        & \qquad &  $\{\epsilon(\bolds{k}) \}= \{\epsilon^{*} (\bolds{k})\}$
\\
        Parity-particle-hole~(${\cal CP}$) symmetry
        & \qquad & ${\cal H}(\bolds{k})= - ({\cal CP}) {\cal H}^{*}(\bolds{k}) ({\cal CP})^{-1}  $ 
        & \qquad & $\{ \epsilon(\bolds{k}) \} = \{- \epsilon^{*}(\bolds{k})\} $
\\
        \hline 
     \end{tabular}
     \vspace{1ex}

     {\raggedright Here the unitary operator $A \in \{ \Gamma, \Lambda, \varsigma, {\cal S}, {\cal P}, {\cal I} \}$ obeys $A^{2}=1$, and the anti-unitary operator $A \in \{ {\cal C}_{\pm}, {\cal T}_{\pm} \}$ satisfies $AA^{*}=\zeta_{A} 1$ with $\zeta_{A}= \pm 1$. Note that the spectra of systems with TRS, PHS, TRS$^{\dagger}$ and PHS$^{\dagger}$ exhibit the Kramers degeneracy~\cite{Esaki2011,Sato2012}. We refer to Appendix~\ref{app:symm_allo_hams} for the specific form of the symmetry-preserving Hamiltonians. \par}
     \label{tab:symm}
\end{table*}

\section{EPs in n-band systems}\label{sec:nEps}

Given a generic $n \times n $ matrix ${\cal H}$, the characteristic polynomial is defined by
\begin{align}
{\cal F}_{\lambda} &= \det[\lambda \id - {\cal H}]
\nonumber \\
&=
\lambda^{n} - \sigma_{1} \lambda^{n-1} + \ldots + (-1)^{n} \sigma_{n} = 0, \label{eq:charac_poly_n}
\end{align}
where 
\begin{align}
\sigma_{1}=\tr[{\cal H}], \qquad \sigma_{n}=\det[{\cal H}],
\end{align}
and other $\sigma_{k}$'s are the sum of $k$th order diagonal minors of $\cal H$. Defining $p_{k}=(-1)^{k} \sigma_{k}$ and $s_{k}=\tr[{\cal H}^{k}]$, we have
\begin{align}
p_{k}= -\frac{
s_{k} + p_{1} s_{k-1} + \ldots + p_{k-1} s_{1}
}{k}, \quad k =1,\ldots n-1.
\end{align}
We can thus express all coefficients of ${\cal F}_{\lambda}$ in terms of $\tr[{\cal H}^{k}]$ and $\det[{\cal H}]$~\cite{Brown1994,Curtright2020}. Having the characteristic polynomial in Eq.~\eqref{eq:charac_poly_n}, one can then calculate its discriminant~${\cal D}[{\cal H}]$. ${\cal D}[{\cal H}]$ is zero when ${\cal F}_{\lambda}$ possesses multiple, say $m$ with $m\leq n$, degenerate roots~$\lambda_{m}$. Those $m$ degenerate roots, whose associated eigenvectors in ${\cal H}$ coalesce, are dubbed $m$th order exceptional points~(EP$m$s) or defective degeneracies. As a result, the Jordan canonical form of ${\cal H}$ with EP$m$s exhibits a Jordan block of dimension $m$ and with eigenvalue $\lambda_{m}$ on the major diagonal.

Adjusting coefficients in Eq.~\eqref{eq:charac_poly_n} can give rise to the appearance of EP$n$s in the eigenspectrum of ${\cal H}$. Subsequently, one can evaluate the number of constraints to observe EP$n$s. More precisely, by setting $\sigma_{1}=\tr[{\cal H}]=0$, which is a trivial shift to the spectrum, we are left with $(n-1)$ complex-valued coefficients, $n-2$ different traces and one determinant. To find EP$n$s, we should thus enforce $2(n-1)$ constraints, i.e., $\Re[\det[{\cal H}]]=0$, $\Im[\det[{\cal H}]]=0$, $\Re[\tr[{\cal H}^k]]=0$ and $\Im[\tr[{\cal H}^k]]=0$ with $k = 2, \ldots, n-1$. We emphasize that EP$n$s occur when all of these $2(n-1)$ constraints are simultaneously enforced. In parameter regimes in which smaller number of constraints are satisfied, lower-order EPs, e.g., EP$m$s with $m \leq n $, may find room to emerge in the spectrum of $n\times n$ dimensional matrices.

An alternative approach to counting the number of constraints in matrices with EP$n$s is based on perturbing ${\cal H}$ close to EP$n$s~\cite{Jiang2020}. Here, we introduce a Jordan block~$J_{n}$ as a description for the EP$n$s in ${\cal H}$ with dimension $n$ and, without loss of generality, diagonal value~$\lambda_{n}=0$. Introducing the perturbation matrix $\delta S$, one can find all insignificant, trivial perturbations using $[\delta S, J_{n}]$~\cite{Jiang2020}. The remaining non-trivial perturbation is a $n \times n $ matrix~$\delta J_n$. The matrix elements of $\delta J_{n}$ are always zero except for $n-1$ complex-valued elements, which are $\delta J_{n, j}$ with $j =1,\ldots n-1$. The summation of the Jordan block and its nontrivial perturbation, namely $J_{n}+ \delta J$, describes the low-energy behavior of ${\cal H}$ close to EP$n$s, which reads
\begin{align}
    {\cal H}_{0}=J_{n}+ \delta J_{n} =
    \begin{pmatrix}
    0 & 1 & 0 & \ldots  & 0  & 0 \\
    0 & 0 & 1 & \ldots  & 0 & 0 \\
    \vdots &    \vdots &  \vdots & \ddots& \vdots & \vdots \\
    0 & 0 & 0 &  \ldots & 0 & 1\\
    \delta J_{n, 1} & \delta J_{n, 2} & \delta J_{n, 3} &  \ldots & \delta J_{n, n-1} & 0
    \end{pmatrix}. \label{eq:Jperturb}
\end{align}
Note that when $\tr[{\cal H}]$ is nonzero, the $(n,n)$ matrix element of $\delta J_{n}$ is also nonzero. 
The matrix elements $\delta J_{n, j}$ are related to the coefficients $\sigma_k$ since the characteristic polynomial of $J_{n}+ \delta J$ is identical to Eq.~\eqref{eq:charac_poly_n}.

In fact, ${\cal H}_{0}$ constructs the (transpose) \emph{Frobenius companion matrix} for the characteristic polynomial in Eq.~\eqref{eq:charac_poly_n}~\cite{Brand1964}, and each of the $\delta J_{n,j}$ is proportional to $\sigma_{n+1-j}$, in particular, $\delta J_{n,j} = (-1)^{n+j} \sigma_{n+1-j}$. This result was also derived in Ref.~\onlinecite{Arnold1971}, and further generalized to describe perturbations of any matrix written in the Jordan normal form.
From this approach, we again realize that $2(n-1)$ constraints are needed to determine the presence of EP$n$s in matrix ${\cal H}$, i.e., $\Re[\delta J_{n,j}] = 0$ and $\Im[\delta J_{n,j}] = 0$ with $j = 1, \ldots, n-1$.

\begin{table*}
     \centering
    \caption{Number of constraints to realize EP$n$s in $n$-band systems}
     \begin{tabular}{l|ll|ll}
        \hline \hline 
        Symmetry  &\qquad & \multicolumn{2}{l}{$\#$ constraints}    \\
        \hline 
        & \qquad & $n \in $ even & $n \in$ odd \\
        \hline
        \hline
         CS 
        & \qquad & $n-1$~
        $
            \begin{cases}
              \Re[\det[{\cal H}]] , \\
             \Re[\tr[{\cal H}^{2l}]] ,\\
             \Im[\tr[{\cal H}^{2l-1}]] .
            \end{cases}
        $
        & $\quad -$
        \\
         \hline 
                psCS
        & \qquad & $n \quad \,\,$ ~ 
       $
            \begin{cases}
              \det[{\cal H}] , \\
              \tr[{\cal H}^{2l}] .
            \end{cases}
      $
        & $n-1$~
        $
            \begin{cases}
              \tr[{\cal H}^{2l}] .
            \end{cases}
        $
        \\
         \hline 
                 SLS 
        & \qquad & $n \quad \,\,$ ~ 
       $
            \begin{cases}
              \det[{\cal H}] , \\
              \tr[{\cal H}^{2l}] .
            \end{cases}
      $
        & $n-1$~
        $
            \begin{cases}
              \tr[{\cal H}^{2l}] .
            \end{cases}
        $
        \\
         \hline 
             
           psH symmetry
        & \qquad & $n-1$~
        $
            \begin{cases}
              \Re[\det[{\cal H}]] , \\
             \Re[\tr[{\cal H}^k]] .
            \end{cases}
        $
        & $n-1$~
       $
            \begin{cases}
              \Re[\det[{\cal H}]], \\
             \Re[\tr[{\cal H}^k]] .
            \end{cases}
        $
        \\
         \hline 
           $\cal PT$ symmetry
        & \qquad & $n-1$~
        $
            \begin{cases}
              \Re[\det[{\cal H}]] , \\
             \Re[\tr[{\cal H}^k]] .
            \end{cases}
        $
        & $n-1$~
       $
            \begin{cases}
              \Re[\det[{\cal H}]] , \\
             \Re[\tr[{\cal H}^k]] .
            \end{cases}
        $
        \\
         \hline 
        $\mathcal{CP}$ symmetry  
        & \qquad & $n-1$~
        $
            \begin{cases}
          \Re[\det[{\cal H}]] , \\
          \Re[\tr[{\cal H}^{2l}]] , \\
          \Im[\tr[{\cal H}^{2l-1}]] .
        \end{cases}
        $
        & $n-1$~
        $
         \begin{cases}
           \Im[\det[{\cal H}]] , \\
           \Re[\tr[{\cal H}^{2l}]] , \\
           \Im[\tr[{\cal H}^{2l-1}]] .      
         \end{cases}
         $
         \\
        \hline 
     \end{tabular}
     \vspace{1ex}

     {\raggedright Here $k\in \{1, \ldots n\}$ and $l \in \{1, \ldots , n/2 \}$. Details are provided in Appendix~\ref{app:epn_symmetry}. Behind the number of constraints we write the specific constraints that need to be satisfied to find EP$n$s. We note that there is no entry for CS with $n \in \textrm{odd}$ as this symmetry is not defined in that case.\par}
     \label{tab:symm_epn}
\end{table*}

From the characteristic polynomial in Eq.~\eqref{eq:charac_poly_n} as well as the perturbed Jordan block in Eq.~\eqref{eq:Jperturb}, we can also deduce how the EP$n$s disperse.
While it is commonly assumed that the series expansion resulting from a perturbation with $\omega$ around an EP$n$, i.e., writing $\mathcal{H}_0 = J_n + \omega \delta J_n$, results in the Puiseux series, $\lambda = \lambda_0 + \sum_{j=1}^\infty \omega^{j/n} \lambda_j$ with $\lambda_1 = \delta J_{n,1}^{1/n}$, this is not generally the case. Indeed, \emph{only} when $\delta J_{n,1} \neq 0$, the Puiseux series is recovered for the energy eigenvalues close to an EP$n$ \cite{Yanyuan1998, Demange2011}. When $\delta J_{n,1} = 0$, the perturbed eigenvalues generally split in different cycles of the form $\lambda = \lambda_0 + \sum_{j=1}^\infty \omega^{j/p} \lambda_j$ with $p<n$, and the different values of $p$ summing up to $n$~\cite{Moro1997, Yanyuan1998, Demange2011}.

Let us now see how this translates into our perturbed Jordan block in Eq.~\eqref{eq:Jperturb}.
In particular, when $\sigma_n \neq 0$ and all other $\sigma_j = 0$ (or equivalently, when $\delta J_{n,1} \neq 0$ and all other $\delta J_{n,j} = 0$), we straightforwardly find that the characteristic polynomial reduces to $\lambda^n + (-1)^n \sigma_n = 0$ (or $\lambda^n  - \delta J_{n,1} = 0$). In this case, the EP$n$ disperses with $e^{2 \pi \i r/n} [(-1)^n \sigma_n]^{1/n}$ ($e^{2 \pi \i r/n} [\delta J_{n,1}]^{1/n}$) for $r = 1, \ldots, n$ \cite{Wilkinson1965, Burke1992}. When $\sigma_{n-1} \neq 0$ and all other $\sigma_j = 0$, (or equivalently, when $\delta J_{n,2} \neq 0$ and all other $\delta J_{n,j} = 0$,) we find $\lambda (\lambda^{n-1} + (-1)^{n-1} \sigma_{n-1}) = 0$ (or $\lambda (\lambda^{n-1}  - \delta J_{n,2}) = 0$) for the characteristic polynomial. Now, the EP$n$ disperses with the $n-1$th root, i.e., $\sim [(-1)^{n-1} \sigma_{n-1}]^{1/(n-1)}$ ($\sim [\delta J_{n,2}]^{1/(n-1)}$), combined with a flat band  with $\lambda = 0$. In general, we thus find that when $\sigma_{j} \neq 0$ ($\delta J_{n,j} \neq 0$) and all other $\sigma_k = 0$ ($\delta J_{n,k} = 0$), the low-energy approximation around the EP$n$ reads $\sim [(-1)^j \sigma_j]^{1/j}$ ($\sim [\delta J_{n,j}]^{1/(n+1-j)}$). When all $\sigma_k \neq 0$ (or all $\delta J_{n,k} \neq 0$), it is no longer possible to find complete analytical solutions for the eigenvalues $\lambda$ when $n \geq 5$. Nevertheless, one can numerically compute explicit solutions for the leading terms~\cite{Moro1997, Lidskii1966}.

So far, we discussed EP$n$s in systems with no additional symmetries. Let us now see how the presence of symmetries affects the appearance of EP$n$s.
Writing the determinant and traces as $\det[{\cal H}] = \prod_i \epsilon_i$ and $\tr[{\cal H}^k] = \sum_i \epsilon_i^k$ with $\epsilon_i$ the eigenvalues of $\cal H$ allows us to make general statements when making use of the energy constraints listed in Table~\ref{tab:symm}. We immediately see that PHS, PHS$^\dagger$, TRS, TRS$^\dagger$, $\mathcal{I}$, and $\mathcal{P}$ symmetry are nonlocal in parameter space as they relate eigenvalues with momentum ${\bf k}$ to eigenvalues with momentum $-{\bf k}$. As such, the presence of these symmetries does not reduce the number of constraints but instead puts a constraint on whether the entries in the Hamiltonian are symmetric or antisymmetric. We thus find that the number of constraints for finding EP$n$s in the presence of these symmetries remains at $2(n-1)$. For the remaining symmetries listed in Table~\ref{tab:symm}, however, there is a reduction in the number of constraints.

In the presence of SLS and psCS, $\{\epsilon({\bf k})\}=\{-\epsilon({\bf k})\}$ dictates that in the case of $n \in \textrm{odd}$, at least one of the eigenvalues is necessarily zero, such that $\det[{\cal H}] = 0$.
For $n \in \textrm{even}$, there is no such argument and we thus generally find $\det[{\cal H}] \neq 0$. Turning to the traces, we see that $\tr[{\cal H}^k] \neq 0$ when $k \in \mathrm{even}$, while $\tr[{\cal H}^k] = 0$ when $k \in \mathrm{odd}$ for all $n$. To find EP$n$s, we thus need to satisfy $n$ constraints when $n \in \textrm{even}$, and $n-1$ constraints when $n \in \textrm{odd}$.
The fact that $\{\epsilon({\bf k})\}=\{-\epsilon({\bf k})\}$ also leads to an interesting consequence when considering the possibility of realizing lower-order EPs in $n$-band systems, namely, the addition of an extra band to an $(n-1)$-band system immediately promotes a possibly existing EP$(n-1)$ to an EP$n$ \emph{as long as} this additional band is coupled to the other bands. As such, there is a notion of fragility in these systems, as also pointed out in Ref.~\onlinecite{Mandal2021} for the case of SLS. However, if a band is added that does not couple to any of the other bands, the EP$(n-1)$ survives even though the energy eigenvalues are $n$-fold degenerate at the EP.

If we instead consider $\mathcal{PT}$ and psH symmetry, we see that $\{\epsilon({\bf k})\}=\{\epsilon^*({\bf k})\}$
implies $\{ \det[{\cal H}], \tr[{\cal H}^k] \} \in \mathbb{R}, \forall  k<n$. This means that we need to satisfy $n-1$ constraints to find an EP$n$ in agreement with what is found in Ref.~\onlinecite{Delplace2021}.
Lastly, considering CS and $\mathcal{CP}$ symmetry, $\{\epsilon({\bf k})\}=\{-\epsilon^*({\bf k})\}$
leads to $\det[{\cal H}] \in \mathbb{R}$ for $n \in \textrm{even}$, $\det[{\cal H}] \in i\mathbb{R}$ for $n \in \textrm{odd}$, $\tr[{\cal H}^k] \in \mathbb{R}, k\in \textrm{even}$ and $\tr[{\cal H}^k] \in i  \mathbb{R}, k\in \textrm{odd}$. This gives us again $n-1$ constraints, which was also found in Ref.~\onlinecite{Delplace2021}, see also Ref.~\onlinecite{footnotechiral}. We summarize the results for SLS, psCS, CS, $\mathcal{PT}$, psH and $\mathcal{CP}$ symmetry in Table~\ref{tab:symm_epn} and refer to Appendix~\ref{app:epn_symmetry} for details on the derivation of these findings.

Now turning back to our results for the dispersion around EP$n$s, we see that in the case of SLS and psCS with $n \in \textrm{odd}$, where $\det[\mathcal{H}]=0$ (i.e., $\sigma_n=0$ or $\delta J_{1,n}=0$), EP$n$s disperse with $\mathcal{O}(\omega^{1/(n-1)})$. Interestingly, this is the only instance of symmetries generically preventing the recovery of the $n$th root dispersion for EP$n$s.

In the following, we explore EP$n$s and the implications of symmetry in greater detail by deriving exact results.
Galois theory~\cite{Everitt2018} implies that characteristic polynomials with dimensions greater than four cannot be expressed as combinations of radicals of rational functions of the polynomial coefficients. Therefore, to present analytical results in terms of radicals, we explore the role of symmetries in modifying the structure and numbers of constraints to detect EP$n$s with $n=2,3,4$.

\begin{table*}
     \centering
    \caption{\label{tab:sympar2}Number of constraints and parameters to realize degenerate points in 2-band systems}
          \begin{tabular}{l|ll|ll|ll}
        \hline \hline 
        Symmetry  &\qquad & Operator &\qquad & $\#$ constraints  &\qquad  & $\#$ parameters \\
        \hline 
        \hline
         No symmetry  
         &\qquad  & -
        &\qquad  & 2~$(\eta,\nu)$
       &\qquad  &  2$\times$3~$(d_{x}, d_{y},d_{z})$
        \\
              PHS with ${\cal C}_{-}{\cal C}_{-}^{*}=1$  
         & \qquad & $\id_{2}$
        & \qquad & 2~$(\eta,\nu)$
        & \qquad & 2$\times$3~$(d_{xIa}, d_{xRa}, d_{yIs}, d_{yRs},d_{zIa}, d_{zRa})$
        \\
          PHS with ${\cal C}_{-}{\cal C}_{-}^{*}=-1$  
         & \qquad & $i \sigma_{y} $
        & \qquad & 2~$(\eta,\nu)$
        & \qquad & 2$\times$3~$(d_{xRs}, d_{xIs},d_{yRs},d_{yIs}, d_{zRs}, d_{zIs})$
        \\
        PHS$^{\dagger}$ with ${\cal T}_{-}{\cal T}_{-}^{*}=1$     
         & \qquad & $\id_{2}  $
        & \qquad & 2~$(\eta,\nu)$     
        & \qquad &  2$\times$3~$(d_{xRa}, d_{xIs} , d_{yRs}, d_{yIa}, d_{zRa}, d_{zIs}) $
        \\
       PHS$^{\dagger}$ with ${\cal T}_{-}{\cal T}_{-}^{*}=-1$     
         & \qquad & $i \sigma_{y}  $
        & \qquad & 2~$(\eta,\nu)$     
        & \qquad &  2$\times$3~$(d_{xRs}, d_{xIa}, d_{yRs},d_{yIa}, d_{zRs} ,d_{zIa})$
        \\
         TRS with ${\cal T}_{+}{\cal T}_{+}^{*}=1$    
         & \qquad & $\id_{2}  $
        & \qquad &  2~$(\eta,\nu)$    
        & \qquad & 2$\times$3~$(d_{xRs}, d_{xIa} , d_{yRa}, d_{yIs}, d_{zRs}, d_{zIa})$  
         \\
     TRS with ${\cal T}_{+}{\cal T}_{+}^{*}=-1$    
         & \qquad & $i \sigma_{y}  $
        & \qquad &  2~$(\eta,\nu)$    
        & \qquad & 2$\times$3~$(d_{xRa},d_{xIs},d_{yRa},d_{yIs},d_{zRa}, d_{zIs})$
         \\
             TRS$^{\dagger}$ with ${\cal C}_{+}{\cal C}_{+}^{*}=1$    
         & \qquad & $\id_{2} $ 
        & \qquad &  2~$(\eta,\nu)$   
        & \qquad &  2$\times$3~ $(d_{xRs}, d_{xIs}, d_{yRa}, d_{yIa}, d_{zRs}, d_{zIs} )$
        \\
        TRS$^{\dagger}$ with ${\cal C}_{+}{\cal C}_{+}^{*}=-1$     
         & \qquad & $i \sigma_{y}  $
        & \qquad &  2~$(\eta,\nu)$   
        & \qquad &  2$\times$3~$(d_{xRa},d_{xIa},d_{yRa},d_{yIa},d_{zRa},d_{zIa})$ 
        \\
        CS 
         & \qquad & $\sigma_{z}$ 
        & \qquad & 1~$(\eta)$
        & \qquad &  3~$(d_{xR}, d_{yR}, d_{zI})$  
        \\
         ${\rm ps  CS}$ 
           & \qquad & $\sigma_{z}$
           & \qquad & 2~$(\eta,\nu)$
           & \qquad & 2$\times$1~$(d_{x})$
          \\
        SLS
         & \qquad & $\sigma_{z}$
        & \qquad & 2~$(\eta,\nu)$
        & \qquad &  2$\times$2~$(d_{x},d_{y})$
        \\
        $\cal I$ symmetry
         & \qquad & $\sigma_{z}$
        & \qquad & 2~$(\eta,\nu)$
        & \qquad & 2$\times$ 3~$( d_{xRs},d_{xIa}, d_{yRs},d_{yIa},d_{zRa},d_{zIs})$     
        \\
        $\rm psH$ 
         & \qquad & $\sigma_{x}$
        & \qquad & 1~$(\eta)$
        & \qquad & 3~$(d_{xR},d_{yI},d_{zI})$
        \\
         $\cal P$ symmetry
          & \qquad & $\sigma_{x}$
        & \qquad & 2~$(\eta,\nu)$
        & \qquad & 2$\times$3~$(d_{xs},d_{ya},d_{za})$
        \\
          $\cal P$ symmetry 
          & \qquad & $\sigma_{z}$
        & \qquad & 2~$(\eta,\nu)$
        & \qquad & 2$\times$3~$(d_{xa},d_{ya},d_{zs})$
        \\
         $\cal PT$ symmetry
          & \qquad & $\sigma_{x}$
         &\qquad  & 1~$(\eta)$  
       &\qquad  &  3~$ (d_{xR},d_{yR},d_{zI})$
        \\
        ${\cal CP}$ symmetry
        & \qquad & $\sigma_{x}$
        & \qquad & 1~$(\eta)$
        & \qquad & 3~$(d_{xI}, d_{yI},d_{zR})$
\\
        \hline 
     \end{tabular}
     \vspace{1ex}

     {\raggedright Here $d_{\cal O}= d_{{\cal O}R} + i d_{{\cal O}I}$ with ${\cal O} \in \{ x,y,z\}$. Symmetric and antisymmetric components of $d_{\cal O}$ with respect to $\bolds{k} \to -\bolds{k}$ are labelled by $d_{{\cal O} \alpha s}$ and $d_{{\cal O}\alpha a}$ with $\alpha \in \{ R,I\}$, respectively. $\eta$ and $\nu$ are introduced in Eq.~\eqref{eq:cond2b}. 
     Note that non-zero parameters might vary by changing the chosen Pauli matrix for each symmetry operator, an example of which is presented for the parity symmetry for which we include two representations. Nevertheless, the number of parameters and constraints remain intact.
    \par}
\end{table*}

\section{EPs in two-band systems}\label{sec:2lvl}
To study second order EPs, we perform a matrix decomposition in the Pauli basis. The most generic two-band Hamiltonian in this representation is given by
\begin{align}
{\cal H}(\bolds{k}) = d_{0}(\bolds{k}) \id_{2} + \bolds{d}(\bolds{k}) \cdot \bolds{\sigma} ,
\label{eq:H2d}
\end{align}
where $\bolds{\sigma}=(\sigma_{x}, \sigma_{y}, \sigma_{z})$ is the vector of Pauli matrices (see Appendix~\ref{app:pauli}), $\id_{2}$ is the $2 \times 2$ identity matrix, $\bolds{k}$ denotes the momentum with the appropriate dimensions, and $ d_{0}$ and $ \bolds{d}= (d_{x}, d_{y}, d_{z}) $ are complex-valued momentum dependent variables. In the following, we drop the momentum dependence for the purpose of brevity, and reinstate it when needed. Considering $\bolds{d} = \bolds{d}_{R} + \i \bolds{d}_{I}$ and ${d}_{0} = {d}_{0R} + \i {d}_{0I}$ with $\{ \bolds{d}_R , \bolds{d}_I \} \in \R$, the eigenvalues cast 
\begin{align}
\lambda_{\pm} &= d_{0} \pm 
\sqrt{ d_{R}^{2} - d_{I}^{2} + 2 \i \bolds{d}_{R}\cdot \bolds{d}_{I}}. \label{eq:eigs_two_band_model}
\end{align}
The characteristic polynomial given in Eq.~\eqref{eq:charac_poly_n} in this case reads
\begin{align}
{\cal F}_{\lambda} (\bolds{k})&=\lambda^2 - \tr[\mathcal{H}]\lambda + \det[\mathcal{H}] = 0, \label{eq:charac_poly_ep2}
\end{align}
such that $\lambda = \left(\tr[\mathcal{H}] \pm \sqrt{\tr[\mathcal{H}]^2 - 4 \, \det[\mathcal{H}] } \right)/2$. Comparing these roots with $\lambda_{\pm}$ given in Eq.~\eqref{eq:eigs_two_band_model}, we get
\begin{align}
    \tr[\mathcal{H}]^2 - 4 \det[\mathcal{H}] = d_{R}^{2} - d_{I}^{2} + 2 \i \bolds{d}_{R}\cdot \bolds{d}_{I}.
\end{align}
The degenerate points are then obtained by setting the discriminant of ${\cal F}_{\lambda} (k)$ in Eq.~\eqref{eq:charac_poly_ep2} to zero, i.e.,
\begin{align}
{\cal D }[{\cal H}] &=\tr[\mathcal{H}]^2 - 4 \det[\mathcal{H}]
=0.\label{eq:disc2b}
\end{align}
The defective degenerate points are the EP2s. Without loss of generality, we can set $\tr[{\cal H}]=2 d_{0}=0$. To find EP2s, we introduce two constraints based on the real~($\eta$) and imaginary~($\nu$) parts of ${\cal D }[{\cal H}]$ as
\begin{align}
\eta=d_{R}^{2} - d_{I}^{2} = 0, \quad \& \quad \nu=\bolds{d}_{R} \cdot \bolds{d}_{I}=0.
\label{eq:cond2b}
\end{align}
Here $\eta$ and $\nu$ describe $N$-spatial-dimensional surfaces.
Note that in Hermitian systems, i.e., $\bolds{d}_{I}=0$, band touching points occur when all components of $\bolds{d}_{R}$ vanish amounting to at most three constraints.

In the vicinity of EP$2$s, the Hamiltonian casts the perturbed Jordan block with dimension $n=2$ in Eq.~\eqref{eq:Jperturb} and reads
\begin{align}
    {\cal H}_{0} = 
    \begin{pmatrix}
        0 & 1 \\
        -\det[{\cal H}(\bolds{k}) ] & 0
    \end{pmatrix}.
\end{align}
Using the similarity transformation ${\cal H}_0 = S \Delta S^{-1}$, the dispersion relation close to the EP$2$s yields $\pm \sqrt{-\det[{\cal H}(\bolds{k}) ]}$, which are the diagonal elements of $\Delta$. This result is central in various studies on systems with $\det[{\cal H}(\bolds{k}) ] =-|\bolds{k}|$ due to the nonanalytical energy dispersion~\cite{Carlstrom2018}.

In two spatial systems, solutions to $\eta = \nu = 0$ [cf. Eq.~\eqref{eq:cond2b}] describe two closed curves in $\bolds{k}$ space, such that EP2s appear when these curves intersect.
In three spatial dimensions, the intersection between the two-dimensional surfaces described by $\eta=0$ and $\nu=0$ forms a closed exceptional curve, which can give rise to exceptional knots and result in exotic features such as open real/imaginary Fermi surfaces~\cite{Carlstrom2018}.

We now turn to the symmetries listed in Table~\ref{tab:symm} and see how the presence of one or the coexistence of multiple symmetries constraints the appearance of EP2s. This problem for EP2s was also studied in Ref.~\onlinecite{Budich2019} for the symmetries defined by Bernard and LeClair~(BLC)~\cite{Bernard2002}~(see also Appendix~\ref{app:symmBL}). We cast it here in the form of the symmetries as given in Table~\ref{tab:symm}, which also includes additional symmetries to the BLC classification. To demonstrate our procedure, we treat two symmetries explicitly as well as their combination in the following.

As an example, we start by considering PHS$^\dagger$ symmetry with $\mathcal{T}_{-} \mathcal{T}_{-}^{*} = -1$. We choose ${\cal T}_{-} = \i \sigma_{y} $, such that the most generic form of a particle-hole~(PH)-symmetric Hamiltonian from Eq.~\eqref{eq:H2d} casts
\begin{align}
{\cal H}_{{\rm PHS}^{\dagger}}
 &= \left( \i d_{0Is} - d_{0Ra}   \right) \id_{2} + \left({\bf d}_{Rs} - \i {\bf d}_{Ia}\right) \cdot \bolds\sigma.
\end{align}
Here we have introduced an additional label onto $d$, where each of the $d$ parameters is represented as $d_{{\cal O}\alpha}=d_{{\cal O}\alpha s}+ d_{{\cal O}\alpha a}$ with ${\cal O} \in \{x,y,z\}$ and $\alpha \in \{R,I\}$ where $d_{{\cal O}\alpha s}~(d_{{\cal O}\alpha a})$ is (anti-)symmetric under $\bolds{k} \to -\bolds{k}$, i.e., $d_{{\cal O}\alpha s}(\bolds{k})=d_{{\cal O}\alpha s}(-\bolds{k})$ and $d_{{\cal O}\alpha a}(\bolds{k})=-d_{{\cal O}\alpha a}(-\bolds{k})$. The trace and determinant then read
\begin{align}
    \tr[\mathcal{H}_{{\rm PHS}^{\dagger}}] &= 2 (\i d_{0Is} -  d_{0Ra}), \\
    \det[\mathcal{H}_{{\rm PHS}^{\dagger}}] &= (\i d_{0Is}-d_{0Ra})^2 
- d_{Rs}^2 + d_{Ia}^2 + 2 \i {\bf d}_{Rs} \cdot {\bf d}_{Ia}.
\label{eq:phsdag_2lvl}
\end{align}
Setting the discriminant in Eq.~\eqref{eq:disc2b} to zero (${\cal D}[\mathcal{H}_{{\rm PHS}^{\dagger}}]=0$), we immediately find modified $\eta,\nu$ constraints, which are
\begin{equation} \label{eq:ep2_phs_dagger_constraints}
   \eta=d_{Rs}^2 - d_{Ia}^2 = 0, \quad \& \quad \nu={\bf d}_{Rs} \cdot {\bf d}_{Ia} = 0.
\end{equation}
The presence of PHS$^\dagger$ thus does not reduce the number of constraints for finding EP2s but merely restricts the momentum-dependency of parameters $d$.

If we instead consider $\mathcal{P}$ symmetry with $\mathcal{P} = \sigma_x$, the most generic form of a parity-symmetric Hamiltonian reads
\begin{align} \label{eq:ep2_ham_parity}
{\cal H}_{\mathcal{P}}
 &= d_{0s} \id_{2} + \left(d_{xs}, - d_{ya}, - d_{za}\right) \cdot \bolds\sigma
.
\end{align}
The trace and determinant then read
\begin{align}
    \tr[\mathcal{H}_{\mathcal{P}}] &= 2 (d_{0Rs}+\i d_{0Is}), \\
    \det[\mathcal{H}_{\mathcal{P}}] &= (d_{0Rs}+\i d_{0Is})^2 
- d_{R \cal P}^2 + d_{I \cal P}^2 + 2 i {\bf d}_{R \cal P} \cdot {\bf d}_{I \cal P}.
\label{eq:phsdag_2lvl}
\end{align}
Here we used ${\bf d}_{R  \cal P} =(d_{xRs}, -d_{yRa}, -d_{zRa}) $ and ${\bf d}_{I  \cal P} =(d_{xIs}, -d_{yIa}, -d_{zIa}) $.
To find EP2s, we satisfy $\eta$ and $\nu$ constraints for this system as
\begin{align}
\begin{cases}
     \eta=d_{xRs}^2 + d_{yRa}^2 + d_{zRa}^2 - d_{xIs}^2 - d_{yIa}^2 - d_{zIa}^2 = 0 , \\
   \nu= d_{xRs}d_{xIs} + d_{yRa}d_{yIa} + d_{zRa}d_{zIa} = 0.  
\end{cases}
\end{align}
Similar to PHS$^\dagger$, $\mathcal{P}$ symmetry puts restrictions on the momentum dependency of the $d_i$'s, while not reducing the number of constraints for realizing EP2s.

If we now consider the presence of both PHS$^\dagger$ with ${\cal T}_{-} = \i \sigma_{y} $ and $\mathcal{P}$ symmetry imposed by $\sigma_{x}$, we get
\begin{align}
{\cal H}_{\cal {\rm \mathcal{P}-PHS}^{\dagger}}
&= \i d_{0Is}\id_{2} + (d_{xRs}, - \i d_{yIa}, - \i d_{zIa})\cdot \bolds{\sigma}.
\end{align}
The trace and determinant then read
\begin{align}
    \tr[\mathcal{H}_{\cal {\rm \mathcal{P}-PHS}^{\dagger}}] &= 2 \i d_{0Is}, \\
    \det[\mathcal{H}_{\cal {\rm \mathcal{P}-PHS}^{\dagger}}] &= -d_{0Is}^2 
- d_{xRs}^2 + d_{yIa}^2 + d_{zIa}^2,
\end{align}
and we find
\begin{align}
   \eta= d_{xRs}^2 - d_{yIa}^2 - d_{zIa}^2 = 0, \qquad \& \quad \nu= 0.
\end{align}
Clearly one merely should satisfy $\eta=0$ to find EP2s in this system. Therefore, even though PHS$^\dagger$ and $\mathcal{P}$ symmetry individually do not reduce the number of constraints, the combination of these symmetries leaves only one constraint nonzero.

We summarize the results for these and the other symmetries in Table~\ref{tab:sympar2}. There we specify the symmetry generator and number of nonvanishing constraints and $d$ parameters in the presence of each symmetry. Table~\ref{tab:2symm_ep2} summarizes various combinations of psH symmetry and other symmetries in the system. We note that one can simply compare the number of parameters in the fourth column of Table~\ref{tab:sympar2} to see which terms survive in the presence of multiple symmetries. As expected, the results in Table~\ref{tab:sympar2} are in agreement with our general findings regarding EP$n$s in Table~\ref{tab:symm_epn}.

\begin{table}
     \centering
    \caption{Summarized combined PsH symmetry with other symmetries and numbers of constraints and parameters}
     \begin{tabular}{l|ll|ll}
        \hline \hline 
        Symmetry  &\qquad & $\#$ constr. &\qquad  & $\#$ parameters \\
        \hline 
        \hline
              psH + CS
        & \qquad & 1~$(\eta)$
        & \qquad &  2~$(d_{xR},d_{zI})$
         \\
                psH + SLS
        & \qquad & 1~$(\eta)$
        & \qquad & 2~$(d_{xR},d_{yI})$
         \\
                psH + $\cal I$
        & \qquad & 1~$(\eta)$
        & \qquad & 3~$(d_{xRs},d_{yIa},d_{zIs})$
         \\
           psH+ PHS with ${\cal C}_{-}{\cal C}_{-}^{*}=1$ 
        & \qquad & 1~$(\eta)$
        & \qquad &  3~$(d_{xRa},d_{yIs},d_{zIa})$
          \\
          psH+ PHS$^{\dagger}$ with ${\cal T}_{-}{\cal T}_{-}^{*}=1$ 
        & \qquad & 1~$(\eta)$
        & \qquad &  3~$(d_{xRa},d_{yIa},d_{zIs})$
        \\
              psH+ TRS with ${\cal T}_{+}{\cal T}_{+}^{*}=1$
        & \qquad & 1~$(\eta)$
        & \qquad &  3~$(d_{xRs}, d_{yIs},d_{zIa})$
          \\
        psH+ TRS$^{\dagger}$ with ${\cal C}_{+}{\cal C}_{+}^{*}=1$
        & \qquad & 1~$(\eta)$
        & \qquad &  3~$(d_{xRs},d_{yIa},d_{zIs})$
          \\
         psH+ PHS with ${\cal C}_{-}{\cal C}_{-}^{*}=-1$ 
        & \qquad &  1~$(\eta)$
        & \qquad &  3~$(d_{xRs}, d_{yIs},d_{zIs} )$
          \\
          psH+ PHS$^{\dagger}$ with ${\cal T}_{-}{\cal T}_{-}^{*}=-1$
        & \qquad &  1~$(\eta)$
        & \qquad &  3~$(d_{xRa},d_{yIs},d_{zIa})$
                \\
  psH+ TRS with ${\cal T}_{+}{\cal T}_{+}^{*}=-1$ 
        & \qquad & 1~$(\eta)$
        & \qquad &  3~$(d_{xRa},d_{yIs},d_{zIs})$
          \\
          psH+ TRS$^{\dagger}$ with ${\cal C}_{+}{\cal C}_{+}^{*}=-1$
        & \qquad & 1~$(\eta)$
        & \qquad &  3~$(d_{xRa},d_{yIa},d_{zIa})$
        \\
        \hline 
     \end{tabular}
     \vspace{1ex}

     {\raggedright One can find nonzero $d$ parameters by keeping common nonzero parameters given by each symmetry individually presented in Table~\ref{tab:sympar2}. While we only list the coexistence of psH symmetry with other non-Hermitian symmetries, the found recipe for determining the number of parameters is generic.
 \par} \label{tab:2symm_ep2}
\end{table}

To demonstrate our findings in this section by a concrete example, we now look at an effective description of the driven-dissipative Kitaev model presented in Ref.~\onlinecite{Sayyad2021}. Here the traceless Hamiltonian is given by
\begin{align}
  {\cal H}_\textrm{ddK} =
  \begin{pmatrix}
      -\i 2 \sqrt{\gamma_{l} \gamma_{g} } & -\i (2 J e^{\i k} + \mu) \\
      \i(2 J e^{-\i k } + \mu) & \i 2 \sqrt{\gamma_{l} \gamma_{g}}
  \end{pmatrix},
\end{align}
where $k$ stands for the momentum index, $J$ is the nearest-neighbor hopping amplitude, $\mu$ denotes the chemical potential, and $\gamma_{l}$ and $\gamma_{g}$ are, respectively, loss and gain coupling rates between the 1D system and the dissipative reservoir. This model displays TRS$^{\dagger}$ with generator $\sigma_{z}$, PHS$^{\dagger}$ with generator $\id$, and CS with generator $\sigma_{z}$~\cite{Sayyad2021}. The trace and determinant of ${\cal H}_\textrm{ddK}$ read
\begin{align}
    \tr[{\cal H}_\textrm{ddK}] &=0,\\
    \det[{\cal H}_\textrm{ddK}] &= 4 \gamma _g \gamma _l-4 J^2-4 J \mu  \cos (k)-\mu ^2.
\end{align}
As a result, the $\eta$ and $\nu$ constraints cast
\begin{align}
    \eta &= 4 \gamma _g \gamma _l-4 J^2-4 J \mu  \cos (k)-\mu ^2 ,\\
    \nu &= 0.
\end{align}
At $k = k_{*}$ in which $\eta=0$, EP2s appear in the spectrum of $ {\cal H}_\textrm{ddK}$. For instance, when $k =0 ~(k=\pi)$, EP2s occur when $2 \sqrt{\gamma_{l} \gamma_{g} } = 2J + \mu ~(2 J - \mu)$ which is consistent with the analysis of Ref.~\onlinecite{Sayyad2021}.

\section{EPs in three-band systems}\label{sec:3lvl}

\begin{table*}
     \centering
    \caption{\label{tab:sympar3}Number of constraints and parameters to realize degenerate points in 3-band systems}
     \begin{tabular}{l|ll|ll|ll}
        \hline \hline 
        Symmetry  &\qquad & Operator &\qquad & $\#$ constr.  &\qquad  & $\#$ parameters \\
        \hline 
        \hline
         No symmetry  
         &\qquad  & -
        &\qquad  &  2$\times$2~$(\eta, \nu)$
       &\qquad  & 2$\times $8~$(
       d_{1},d_{2},d_{3},d_{4},d_{5},d_{6},d_{7},d_{8})$
        \\
        PHS with ${\cal C}_{-}{\cal C}_{-}^{*}=1$  
         & \qquad & $\id_{3}$
        & \qquad & 2$\times$2~$(\eta, \nu)$
        & \qquad & 2$\times $8~\makecell{$(d_{1Rs}, d_{2Rs}, d_{3Rs}, d_{4Ra}, d_{5Ra}, d_{6Ra}, d_{7Ra},d_{8Ra}$\\$ d_{1Is}, d_{2Is}, d_{3Is}, d_{4Ia}, d_{5Ia}, d_{6Ia}, d_{7Ia},d_{8Ia})$}
        \\
       PHS$^{\dagger}$ with ${\cal T}_{-}{\cal T}_{-}^{*}=1$     
         & \qquad & $\id_{3}$
        & \qquad & 2$\times$2~$(\eta, \nu)$
        & \qquad & 2$\times $8~\makecell{$(d_{1Rs}, d_{2Rs}, d_{3Rs}, d_{4Ra}, d_{5Ra}, d_{6Ra}, d_{7Ra},d_{8Ra}$\\$d_{1Ia}, d_{2Ia}, d_{3Ia}, d_{4Is}, d_{5Is}, d_{6Is}, d_{7Is},d_{8Is})$}
        \\
       TRS with ${\cal T}_{+}{\cal T}_{+}^{*}=1$    
         & \qquad & $\id_{3}$
        & \qquad & 2$\times$2~$(\eta, \nu)$
        & \qquad & 2$\times $8~\makecell{$(d_{1Ra}, d_{2Ra}, d_{3Ra}, d_{4Rs}, d_{5Rs}, ,d_{6Rs}, d_{7Rs},d_{8Rs}$\\$d_{1Is}, d_{2Is}, d_{3Is}, d_{4Ia}, d_{5Ia}, d_{6Ia}, d_{7Ia} , d_{8Ia})$ }
         \\
       TRS$^{\dagger}$ with ${\cal C}_{+}{\cal C}_{+}^{*}=1$    
         & \qquad & $\id_{3}$
        & \qquad & 2$\times$2~$(\eta, \nu)$
        & \qquad & 2$\times $8~\makecell{$ d_{1Ra},  d_{2Ra},d_{3Ra},  d_{4Rs}, d_{5Rs},  d_{6Rs}, d_{7Rs}, d_{8Rs}$\\$ d_{1Ia},d_{2Ia},  d_{3Ia}, d_{4Is},  d_{5Is},d_{6Is}, d_{7Is},  d_{8Is})$}
        \\
                 ${\rm ps  CS}$ 
           & \qquad & $\frac{\id_{3}}{3} + M^{7} -\frac{M^{8} }{\sqrt{3} }$
           & \qquad & 2~$\times$1~$(\eta)$
           & \qquad & 6~$(d_{2R}, d_{2I},d_{4R},d_{4I},d_{6R},d_{6I})$
          \\
         SLS
         & \qquad & $\frac{\id_{3}}{3} + M^{7} -\frac{M^{8} }{\sqrt{3} }$
        & \qquad & 2~$\times$1~$(\eta)$
        & \qquad & 8~$(d_{1R}, d_{1I}, d_{3R},d_{3I},d_{4R},d_{4I},d_{6R},d_{6I})$
        \\
       $\cal I$ symmetry
         & \qquad & $\frac{\id_{3}}{3} +M^{6} +\frac{M^{7}}{2} + \frac{M^{8}}{2 \sqrt{3}}$
        & \qquad & 2$\times$2~$(\eta, \nu)$
        & \qquad & 2$\times$8~$(d_{1},d_{2},d_{3Ra}, d_{3Is}, d_{4},d_{5},d_{6Rs},d_{6Ia}, d_{7},d_{8})$
        \\
       $\rm psH$ 
         & \qquad & $\frac{\id_{3}}{3} + M^{4} -\frac{M^{8} }{\sqrt{3} }$
        & \qquad & 2~$(\eta_{R}, \nu_{R})$
        & \qquad & 12~$(
         d_{1I}, d_{2R},d_{2I}, d_{3R}, d_{3I}, d_{4R}, d_{5R},d_{5I}, d_{6R},  d_{6I}, d_{7I},d_{8R})$
        \\
        $\cal P$ symmetry
         & \qquad & $\i \frac{\id_{3}}{3} + M^{7} -\i \frac{M^{8} }{\sqrt{3} }$
        & \qquad &  2$\times$2~$(\eta, \nu)$
        & \qquad & 2$\times$8~\makecell{$(d_{1Ra},d_{2Rs}, d_{3Ra},d_{4Ra}, d_{5Rs},d_{6Ra},d_{7Rs}, d_{8Rs}$\\$d_{1Ia},d_{2Is},d_{3Ia},  d_{4Ia},d_{5Is},d_{6Ia} ,d_{7Is}, d_{8Is})$}
        \\
         $\cal PT$ symmetry
         & \qquad & $\i \frac{\id_{3}}{3} + M^{7} -\i \frac{M^{8} }{\sqrt{3} }$
        & \qquad &  2~$(\eta_{R}, \nu_{R})$
        & \qquad & 12~$(
        d_{1R}, d_{2R}, d_{2I},d_{3R},d_{3I} ,d_{4I},d_{5R},d_{5I}, d_{6R}, d_{6I}, d_{7R},d_{8R})$
        \\
        $\cal PT$ symmetry
         & \qquad & $ \frac{\id_{3}}{3} + M^{7} - \frac{M^{8} }{\sqrt{3} }$
        & \qquad &  2~$(\eta_{R}, \nu_{R})$
        & \qquad & 8~$(
        d_{1R},
        d_{2I},
        d_{3R},
         d_{4I}, 
          d_{5R},  
          d_{6I},
        d_{7}, 
        d_{8}
        )$
        \\
        ${\cal CP}$ symmetry
        & \qquad & $\i \frac{\id_{3}}{3} + M^{7} -\i \frac{M^{8} }{\sqrt{3} }$
        & \qquad & 2~$(\eta_{R}, \nu_{I})$
        & \qquad & 8~$
        (d_{1I}, d_{2R}, d_{3I}, d_{4R}, d_{5I}, d_{6R}, d_{7I}, d_{8I})
        $
\\
        \hline 
     \end{tabular}
     \vspace{1ex}

     {\raggedright Here $d_{\cal O}= d_{{\cal O}R} + \i d_{{\cal O}I}$ with ${\cal O} \in \{ x,y,z\}$. Symmetric and antisymmetric components of $d_{\cal O}$ with respect to $\bolds{k} \to -\bolds{k}$ are labelled by $d_{{\cal O}s}$ and $d_{{\cal O}s}$, respectively. Complex-valued $\eta$ and $\nu$ are introduced in Eqs.~(\ref{eq:eta3lvl}, \ref{eq:nu3lvl}). $\nu_{R},\eta_{R}$ stand for the real components of $\nu,\eta$. Note that nonzero parameters might vary by changing the depicted symmetry operators for each symmetry, see, e.g., the two different choices for $\cal PT$ symmetry. Nevertheless, the number of parameters and constraints remain intact. \par}
\end{table*}

To study EPs of order three, we perform a matrix decomposition in the Gell-Mann basis.
Within this decomposition, the most generic three-band Hamiltonian is given by
\begin{align}
{\cal H}(\bolds{k}) = d_{0}(\bolds{k}) \id_{3} +\bolds{d}(\bolds{k}) \cdot {\bf M}  ,
\label{eq:H3d}
\end{align}
where ${\bf M}=(M_1, M_2, \ldots, M_8)$ is the vector of traceless three-band Gell-Mann matrices~(see Appendix~\ref{app:gellmann_su3}), $\id_{3}$ is the $3 \times 3$ identity matrix, $\bolds{k}$ denotes the momentum with the appropriate dimensions, and $( d_{0}(\bolds{k}), \bolds{d}(\bolds{k})  )$ are complex-valued momentum dependent variables.

For the $3 \times 3$ matrix ${\cal H}$ in Eq.~\eqref{eq:H3d}, the characteristic polynomial in Eq.~\eqref{eq:charac_poly_n} reads
\begin{align}
{\cal F}_{\lambda}= \lambda^{3} - \tr[{\cal H}] \lambda^{2} + \frac{(\tr[{\cal H}])^{2} - \tr[{\cal H}^2]}{2} \lambda - \det[{\cal H}]=0
.
\label{eq:char_poly_3}
\end{align}
The three solutions $\lambda_{1}, \lambda_{2}, \lambda_{3}$ of ${\cal F}_{\lambda}$ are eigenvalues of $\cal H$ in Eq.~\eqref{eq:H3d} and are given explicitly in Appendix~\ref{app:eigs_three_band}. The associated discriminant for Eq.~\eqref{eq:char_poly_3} then casts
\begin{align}
{\cal D}
 &= - \frac{1}{27}[4 \eta^{3} + \nu^{2}].
\end{align}
Here, the complex-valued constraints read
\begin{align}
\eta &= \frac{\tr[{\cal H}]^2}{2}-\frac{3 \tr[{\cal H}^2]}{2} ,
\label{eq:eta3lvl}\\
\nu &= 27 \det[{\cal H}]-\frac{5 \tr[{\cal H}]^3}{2}+\frac{9 \tr[{\cal H}] \tr[{\cal H}^2]}{2}.
\label{eq:nu3lvl}
\end{align}

In the presence of symmetries, the number of nonzero constraints may reduce and different $d$'s may vanish. Table~\ref{tab:sympar3} summarizes these constraints and the number of nonzero parameters in Hamiltonians with a specific symmetry, listed in Table~\ref{tab:symm}. As before, although we depict a particular symmetry generator for each symmetry, the number of constraints and nonzero parameters do not depend on our choice of generator. This can be explicitly seen for $\cal PT$ symmetry, for which we have presented two possible symmetry operators. Similar to the case of EP2s, three-band touchings occur in the Hermitian case ($\bolds{d}_I = 0$) when $\bolds{d}_R = 0$ amounting to at most eight constraints.
To find EP3s, however, we may again set $\tr[{\cal H}]=0$ without loss of generality, and satisfy four real constraints, $\Re[\det[{\cal H}]] = 0$, $\Im[\det[{\cal H}]] = 0$, $\Re[\tr[{\cal H}^2]] = 0$ and $\Im[\tr[{\cal H}^2]] = 0$, or equivalently $\Re[\eta] = 0$, $\Im[\eta] = 0$, $\Re[\nu] = 0$ and $\Im[\nu] = 0$.

Perturbing close to an EP3 gives [cf. Eq.~\eqref{eq:Jperturb}]
\begin{align}
{\cal H}_{0}= 
\begin{pmatrix}
0 & 1 & 0  \\
0 & 0 & 1  \\
\det[{\cal H}] & \frac{ \tr[{\cal H}^2]}{2} & 0
\end{pmatrix}.
\end{align}
From this, we see that depending on the values of $\det[\cal H]$ and $\tr[{\cal H}^2]$, it is possible to realize different types of EP3s. Note that, without loss of generality, we again set $\tr[{\cal H}]$ to zero.

For systems in which $\tr[{\cal H}^2]\neq0$ and $\det[{\cal H}]=0$, the Jordan decomposition of ${\cal H}_0$ reveals EPs whose low-energy bands consist of a flat band with energy $0$ and two bands with dispersion $ \pm \sqrt{ \tr[{\cal H}^2] /2}$. In Section~\ref{sec:nEps}, we showed that $\det[\cal H]$ is always zero for systems with odd $n$ in the presence of SLS and psCS. In the presence of these symmetries, one can thus only find this type of EP3. An explicit example of this type of EP3 is reported in a system with SLS symmetry in Ref.~\onlinecite{Mandal2021}, see also Ref.~\onlinecite{parity_sls}.

For Hamiltonians in which by construction $\tr[{\cal H}^2]=0$ and $\det[{\cal H}] \neq 0$, the Jordan decomposition of ${\cal H}_0$ suggests it is possible to get a second type of EP3, whose low-energy dispersion yields $ (-1)^{j+j/3} \sqrt[3]{ \det[{\cal H}]}$ for $j =1,2,3$. 

A third type of EP3s can emerge when the constraints $\nu$ and $\eta$, respectively, in Eqs.~(\ref{eq:eta3lvl},\ref{eq:nu3lvl}) are purely real,  i.e., $\Im[\nu]=\Im[\det[{\cal H}]]=0$ and $\Im[\eta]=\Im[\tr[{\cal H}^{2}]]=0$. In this case, the low-energy dispersion can be obtained from the generic solution of $\lambda_{j}$ with $j=1,2,3$ in Eqs.~(\ref{eq:lam1-lvl3}, \ref{eq:lam2-lvl3}, \ref{eq:lam3-lvl3}). If $\Re[\eta]\propto\Re[\tr[{\cal H}^{2}]]$ decrease to zero faster than $\Re[\nu]\propto \Re[\det[{\cal H}]]$ close to EP3, the dominant terms in the low-energy dispersion should be proportional to $\{ \sqrt[3]{\Re[\nu]}, (\i +\sqrt{3})\sqrt[3]{\Re[\nu]}, (\i -\sqrt{3})\sqrt[3]{\Re[\nu]} \}$. This type of EP3 is explicitly studied in a $\cal PT$-symmetric Hamiltonian in Ref.~\onlinecite{Mandal2021}.

We summarize these three types of EP3s in Table~\ref{tab:Ep3s}. Without reference to symmetries, types I and III were also reported in Ref.~\onlinecite{Demange2011}. 
\begin{table}
     \centering
    \caption{\label{tab:Ep3s}Various possibilities of EP3s and their energy dispersion}
     \begin{tabular}{l|ll|ll}
     \hline \hline 
           & \quad& Condition  & \quad& Energy dispersion \\
        \hline 
        \hline
        EP3 0 \quad&  & $\eta\neq0, \nu\neq0$ & \quad&  $(\lambda_{1}, \lambda_{2}, \lambda_{3})$ \\
           &\quad &  &\quad & \\
        EP3 I \quad&  & $\det[{\cal H}]=0$ & \quad&  $ 0,\pm \sqrt{\frac{ \tr[{\cal H}^2]}{2}}$ \\
           &\quad &  &\quad & \\
        EP3 II \quad&  & $\tr[{\cal H}^{2}]=0$ &\quad&  $ (-1)^{j+j/3} \sqrt[3]{ \det[{\cal H}]}$ \\
        &\quad &  &\quad & \\
        EP3 III\quad&  & $\Im[\eta]=\Im[\nu]=0$ &\quad& 
       $ \sqrt[3]{\Re[\nu]}, \alpha \sqrt[3]{\Re[\nu]}, \alpha^{*} \sqrt[3]{\Re[\nu]} $
        \\
         \hline
     \end{tabular}
     
     {\raggedright Here $j \in \{1,2,3\}$ and $\alpha = (\i +\sqrt{3})$. 
    Note that in all cases \emph{four} real constraints should be satisfied to observe EP3s.
    These constraints are counted by two complex equations either $(\eta=0,\nu=0)$ or $(\tr[{\cal H}^{2}]=0, \det[{\cal H}]=0)$. $(\lambda_{1}, \lambda_{2}, \lambda_{3})$ are given in Eqs.~(\ref{eq:lam1-lvl3}, \ref{eq:lam2-lvl3}, \ref{eq:lam3-lvl3}). For EP3 III, $\Re[\eta]$ goes faster to zero than $\Re[\nu]$. See details in the main text. \par}
\end{table}

Aside from EP3s, three-band systems may also host EP2s. To explore the conditions in which these EPs can be realized, we introduce a subclass of traceless $3\times 3$ Hamiltonians, which read
\begin{align}
{\cal H}_{1} &=
\begin{pmatrix}
-(b+e) & 0 &0 \\
0 & b & c \\
0 & d & e
\end{pmatrix} = 
\begin{pmatrix}
-(b+e) & 0_{1\times 2} \\
0_{2 \times 1} & h_{2 \times 2}
\end{pmatrix} ,
\label{eq:H3lvl_2}
\\
&=
h_{3} M^{3}
+ h_{6} M^{6}
+h_{7} M^{7}
+h_{8} M^{8},
\end{align}
where $b,c,d,e$ are complex values, and $a=-(b+e)$, $h_{3}= \i (c-d) /2 $, $h_{6}=(c+d)/2 $, $h_{7}=(a-b)/2$, and $h_{8}=(a+b-2e)/ (2 \sqrt{3}) $. The associated characteristic polynomial for ${\cal H}_1$ in Eq.~\eqref{eq:H3lvl_2} is
\begin{align}
 (b+e+\lambda ) \left(b e-b \lambda -c d-e \lambda +\lambda ^2\right) =0
 ,\label{eq:factorized_char_2ep_in3band}
\end{align}
where the second factor originates from $h_{2 \times 2}$. For this factor, we can write the companion matrix
\begin{align}
    h_{2 \times 2 }
    =
    \begin{pmatrix}
    0 & 1 \\
    -\det[h_{ 2 \times 2}] & \tr[h_{2 \times 2}]
    \end{pmatrix},
\end{align}
which explicitly shows the possibility of observing EP2s in this subsystem of three-band Hamiltonians.

\begin{figure}
    \centering
    \includegraphics[width=0.98\columnwidth]{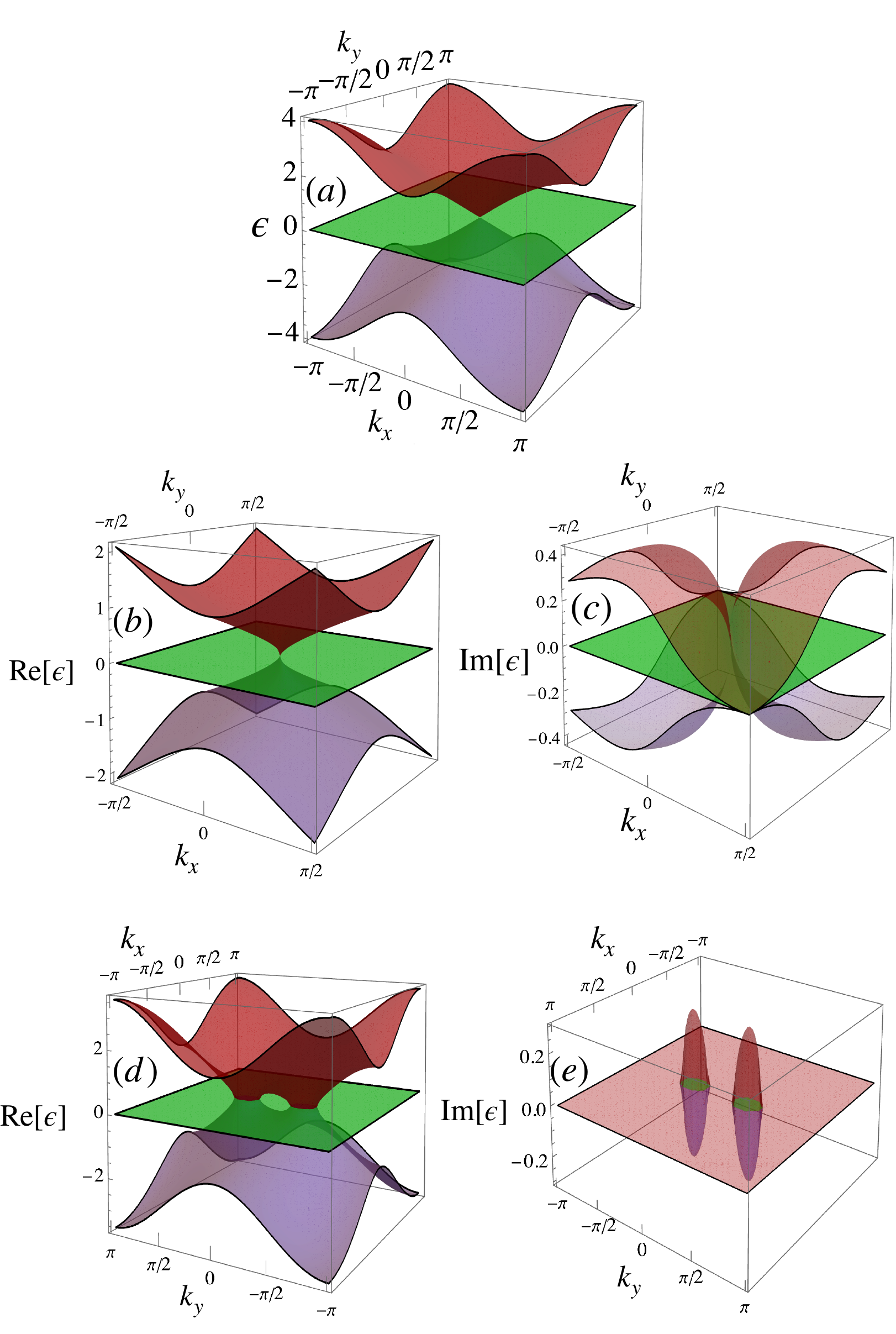}
    \caption{(a) The spectrum of the three-band model in Eq.~\eqref{eq:Hex13band} in its Hermitian limit with $\alpha_{x}=\alpha_{y}=\alpha_{z}=0$. (Middle panels) The real~(b) and imaginary~(c) components of the band structure for the non-Hermitian model in Eq.~\eqref{eq:Hex13band} in the presence of psCS with $\alpha_{x}=\alpha_{y}=0.3$ and $\alpha_{z}= \i \sqrt{0.6}$.
    (Bottom panels) The real~(d) and imaginary~(e) components of the the band structure for the non-Hermitian model in the presence of psCS and ${\cal PT}$ symmetry given in Eq.~\eqref{eq:Hex13band_pt} with $\alpha_{x}=\alpha_{y}=0.3$ and $\alpha_{z}= \i \sqrt{0.6}$. Line colors in middle and bottom panels are chosen such that largest~(smallest) values are presented in red~(blue). Smaller ranges for $k_{x},k_{y}$ are for a better visibility purpose.
    }
    \label{Fig:Hex13band}
\end{figure}

\begin{figure}
    \centering
    \includegraphics[width=0.85\columnwidth]{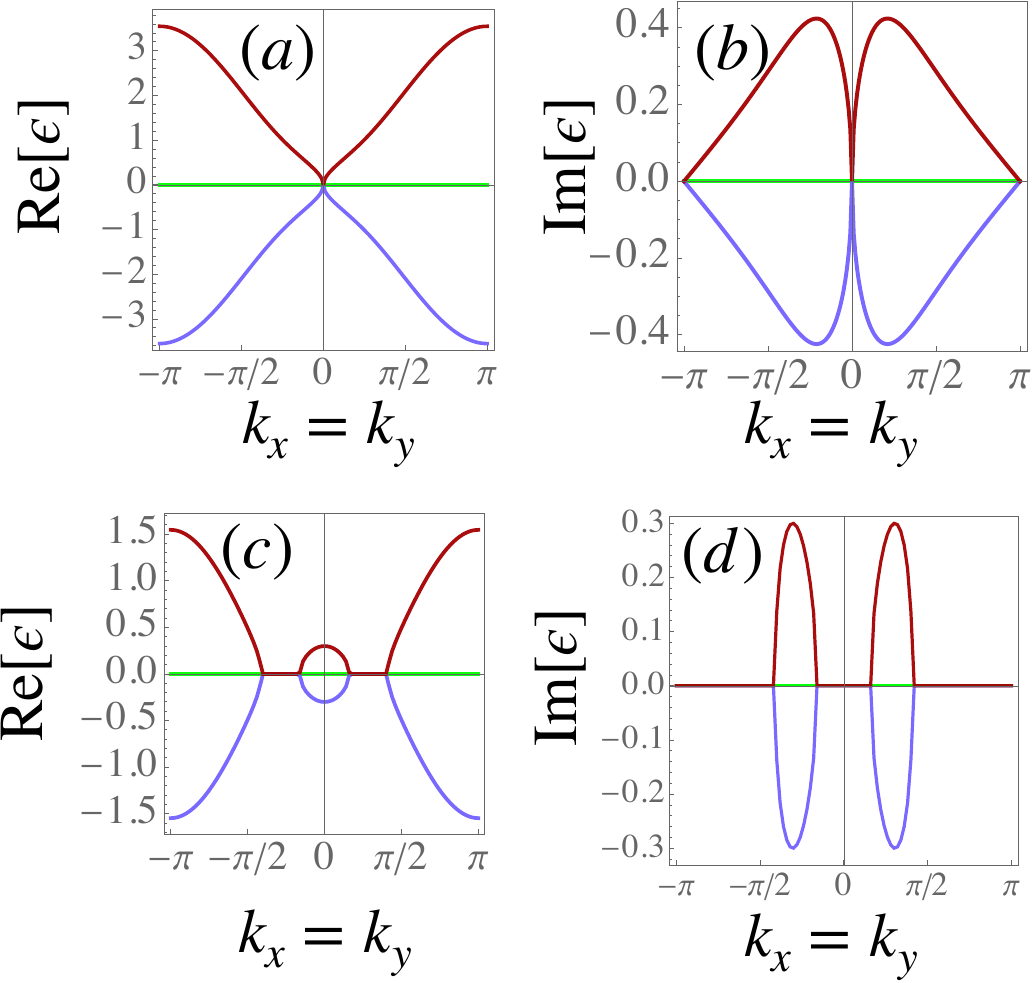}
    \caption{ (Upper panels) The same as panels (b,c) in Fig.~\ref{Fig:Hex13band} along the $k_{x}=k_{y}$ direction.
    (Bottom panels) The same as panels (d,e) in Fig.~\ref{Fig:Hex13band} along the $k_{x}=k_{y}$ direction.
    }
    \label{Fig:Hex13band_cut}
\end{figure}

To explore the effect of imposing symmetries on the behavior of EPs and the associated conditions for their appearance, we introduce an explicit three-band model in the following.
Our model Hamiltonian reads
\begin{align}
{\cal H} =
\left(
\begin{array}{ccc}
 0 
 & h_{x} 
 & -h_{y}
 \\
 -h_{x}
 & 0 
 & h_{z}
 \\
 h_{y} 
 & -h_{z}
 & 0 \\
\end{array}
\right)
,\label{eq:Hex13band}
\end{align}
where $h_{x}=\alpha_{x}+\i \sin (k_{x}) $, $h_{y}= \alpha_{y}+\i \sin (k_{y}) $, and $h_{z}=\alpha_{z} + \i (-2 + \cos(k_{x}) + \cos(k_{y}) ) $. Our model is a non-Hermitian generalization of the effective Hamiltonian for three-fold fermions at $k_{z}=\pi/2$ introduced in Ref.~\onlinecite{Bradlyn2016}. This model Hamiltonian displays the pseudo-chiral symmetry with generator $-\id_{3}$ and hosts a threefold degeneracy in its Hermitian spectrum at $\alpha_{x}=\alpha_{y}=\alpha_{z}=0$, as shown in Fig.~\ref{Fig:Hex13band}~(a). The traces and the determinant of this model read
\begin{align}
    \tr[{\cal H}] &=0,\\
    -\frac{\tr[{\cal H}^2]}{2} &=  
    \alpha_{x}^2+\alpha_{y}^2+\alpha_{z} (\alpha_{z}-4 \i)
    +2 \i \alpha_{x} \sin (k_{x})
    \nonumber \\
    &
    +\cos (k_{x}) (2 \i \alpha_{z}-2 \cos (k_{y})+4)+2 \i \alpha_{y} \sin (k_{y})
    \nonumber \\
    &
    +(4+2 \i \alpha_{z}) \cos (k_{y})-6
 ,\\
    \det[{\cal H} ] &=
    0 .
\end{align}
For the purpose of simplicity, we set $\alpha_{x}=\alpha$, $\alpha_{y}=\alpha$ and $\alpha_{z}=\i \sqrt{2 \alpha^{2}}$ with $\alpha$ a real-valued number.
The real and imaginary parts of the eigenvalues with $\alpha=0.3$ are shown in Figs.~\ref{Fig:Hex13band}(b,c) and Figs.~\ref{Fig:Hex13band_cut}(a,b), respectively, and reveal that our system exhibits an EP3 when $(k_x,k_y) \to 0$.
Based on the low-energy dispersion of the spectrum with one flat~(with energy zero) and two dispersive bands, Table~\ref{tab:Ep3s} suggests that we are dealing with an EP3 I. To examine this suggestion, we look at the band structure of our model at small momenta~($\{k_{x},k_{y}\} \to 0$)
\begin{align}
  \epsilon_{1} &= 0,\\
  \epsilon_{2}&= -\i  \sqrt{-k_{x}^2+2 \i \alpha  (k_{x}+k_{y})-k_{y}^2},\\
  \epsilon_{3}&=\i \sqrt{-k_{x}^2+2 \i \alpha  (k_{x}+k_{y})-k_{y}^2}.
\end{align}
The factor which is under the square root in $\epsilon_{2}$ and $\epsilon_{3}$ is $-\tr[{\cal H}^{2}]/2$. Thus, our model in Eq.~\eqref{eq:Hex13band} indeed gives rise to type I EP3s.  

Imposing ${\cal PT}$ symmetry with generator $\id_{3}/3 + M^{7} - M^{8}/\sqrt{3}= {\rm diag}(1,-1,1)$ on this model leads to a pseudo-chiral-${\cal PT}$-symmetric Hamiltonian, which reads
\begin{align}
    {\cal H}_{\cal PT}
    &=
\left(
\begin{array}{ccc}
 0 & \i \sin (k_{x}) & -\alpha  \\
 -\i \sin (k_{x}) & 0 & \i h_{\alpha} \\
 \alpha  & -\i h_{\alpha} & 0 \\
\end{array}
\right),
\label{eq:Hex13band_pt}
\end{align}
where $h_{\alpha}=\left[\sqrt{2} \alpha +\cos (k_{x})+\cos (k_{y})-2\right]$.
The band structure of this system at $\alpha=0.3$ is plotted in Figs.~\ref{Fig:Hex13band}~(d,e) and Fig.~\ref{Fig:Hex13band_cut}~(c,d). Even though we observe three-band crossings in the band structure of this system, we emphasize that EP2s, instead of EP3s, emerge at momenta slightly away from the origin. To demonstrate this statement, we look at the characteristic polynomial, which reads
\begin{align}
    -\lambda  \left[\lambda ^2 -\Omega_{\alpha}(k_x, k_y) \right]
    =0,\label{eq:char_pol_Hpt_ex3band}
\end{align}
where $\Omega_{\alpha}(k_{x}, k_{y})=-\alpha^2+4 \sqrt{2} \alpha -2 \sqrt{2} \alpha  \cos (k_{x})-2 \cos (k_{x}) \cos (k_{y})-\sin ^2(k_{x})-\cos ^2(k_{x})+4 \cos (k_{x})-2 \sqrt{2} \alpha  \cos (k_{y})-\cos ^2(k_{y})+4 \cos (k_{y})-4$. The characteristic polynomial in Eq.~\eqref{eq:char_pol_Hpt_ex3band} factorizes into a first order and a second-order polynomial, similar to Eq.~\eqref{eq:factorized_char_2ep_in3band}. This means that it is possible to find a unitary transformation such that the Hamiltonian matrix in Eq.~\eqref{eq:char_pol_Hpt_ex3band} features a zero row and zero column, or in other words, that the zero-energy flat band is not coupled to the other bands. We also note that $\Omega_{\alpha}(k_{x},k_{y})=0$ delineates the region in which EP2s exist.

Lastly, we consider the case of finding EP2s in a three-band model in the presence of SLS. We start with the Hamiltonian in Eq.~\eqref{eq:H3lvl_2}, which is SL-symmetric with $S = \id_{3}/3 + M^7 - M^8/\sqrt{3}={\rm diag}(1,-1,1)$ as defined in Table~\ref{tab:sympar3} when $e = -b$, such that ${\cal H}_{1, \textrm{SLS}} = h_3 M^3 + h_6 M^6$. The eigenvalues for this Hamiltonian read $0, \pm \sqrt{b^2 - cd}$. Even though the eigenvalues are three-fold degenerate when $b^2 = - cd$, only two eigenvectors coalesce onto one at this point, and we thus find an EP2 in the system. There are two important things to note for this example. Firstly, it is only possible to find EP2s in three-band models with SLS \emph{as long as} the zero-energy band is not coupled to the other bands, or in other words, \emph{as long as} the three-band model can be described by a Hamiltonian like Eq.~\eqref{eq:H3lvl_2}. Indeed, if the zero-energy band were to be coupled to the other bands such that the most generic three-band SL-symmetric Hamiltonian reads ${\cal H}_\textrm{SLS} = d_1 M^1 + d_3 M^3 + d_4 M^4 + d_6 M^6$ [cf. Table~\ref{tab:sympar3}], any previously existing EP2 would immediately be promoted to an EP3. Secondly, to retrieve ${\cal H}_{1, \textrm{SLS}}$, one has to tune $d_1 = d_4 = 0$. This means that to find an EP2 in a three-band SL-symmetric model, one has to satisfy six real constraints, namely the two constraints, $\Re[b^2] = - \Re[cd], \Im[b^2] = - \Im[cd]$, that one needs to satisfy to find an EP2 in the presence of SLS (cf. Table~\ref{tab:sympar2}) as well as the additional four constraints 
$\Re[d_1] = \Im[d_1] = \Re[d_4] = \Im[d_4] = 0$.

\section{EPs in four-band systems}\label{sec:4lvl}

\begin{table*}
     \centering
    \caption{\label{tab:sympar4I}Number of constraints and parameters to realize degenerate points in 4-band systems}
     \begin{tabular}{l|ll|ll|ll}
        \hline \hline 
        Symmetry  &\quad & Operator &\quad & $\#$ constr.  &\quad  & $\#$ parameters \\
        \hline 
        \hline
         No symmetry  
         &\qquad  & -
        &\qquad  & \makecell{2$\times$3
        \\
        $(\nu, \eta, \kappa)$
        }
       &\qquad  & 2$\times$15~$(d_{1} , d_{2}, d_{3}, d_{4}, d_{5}, d_{6}
       d_{7}, d_{8}, d_{9}, d_{10}, d_{11}, d_{12}, d_{13}, d_{14}, d_{15})$
        \\
         \hline 
        PHS with ${\cal C}_{-}{\cal C}_{-}^{*}=1$  
         & \qquad & $\Lambda^{8} + \Lambda^{11}$
        & \qquad & \makecell{2$\times$3
        \\
        $(\nu, \eta, \kappa)$
        }
        & \qquad & 2$\times$15~
        \makecell{$
(d_{1Ra},d_{1Rs},
d_{2Ra},d_{3Ra},d_{3Rs},
d_{4Ra},d_{4Rs},d_{5Ra},
d_{6Ra},d_{6Rs},
d_{7Ra},       $\\$ 
d_{7Rs},
d_{8Ra},
d_{9Ra},d_{9Rs},d_{10Ra},d_{10Rs},
d_{11Ra}, d_{12Ra},d_{12Rs},
   $\\$ 
d_{13Ra},d_{13Rs},
d_{14Ra},d_{14Rs},
d_{15Rs},
 d_{15Ra},
 d_{1Ia}, d_{1Is},
d_{2Ia},
d_{3Ia},d_{3Is},
       $\\$ 
d_{4Ia},d_{4Is},
d_{5Ia}, d_{6Ia},d_{6Is},
d_{7Ia},d_{7Is}, d_{8Ia},
d_{9Ia},d_{9Is},d_{10Ia},d_{10Is},
       $\\$ 
d_{11Ia},
 d_{12Ia},d_{12Is},
d_{13Ia},d_{13Is},d_{14Ia},d_{14Is},
d_{15Ia},d_{15Is}
)   
        $
        }
        \\
         \hline 
        PHS with ${\cal C}_{-}{\cal C}_{-}^{*}=-1$  
         & \qquad & $-\i (\Lambda^{1} + \Lambda^{6})$
        & \qquad & \makecell{2$\times$3
        \\
        $(\nu, \eta, \kappa)$
        }
        & \qquad & 2$\times$15~\makecell{$(
      d_{1Rs},d_{2Rs},d_{2Ra}, 
      d_{3Ra},d_{3Rs},d_{4Ra},d_{4Rs},
      d_{5Ra},d_{5Rs},d_{6Rs},
             $\\$ 
      d_{7Rs}, d_{8Ra},d_{8Rs},      
       d_{9Ra},d_{9Rs},
         d_{10Ra},d_{10Rs},
       d_{11Ra},d_{11Rs},
       $\\$        
       d_{12Rs},
       d_{13Rs}, d_{14Ra},d_{14Rs},
       d_{15Rs},d_{15Ra},  
      d_{1Is}, d_{2Ia},
       $\\$      
      d_{2Is},
      d_{3Ia},d_{3Is},d_{4Ia},d_{4Is},
      d_{5Ia},d_{5Is},
       d_{6Is},d_{7Is}, 
       d_{8Ia},d_{8Is},
       d_{9Ia},d_{9Is},
        $\\$
       d_{10Ia},d_{10Is},d_{11Ia},d_{11Is},
       d_{12Is}, d_{13Is},
     d_{14Is},d_{14Ia},d_{15Ia},d_{15Is}
        )$}
        \\
         \hline 
       PHS$^{\dagger}$ with ${\cal T}_{-}{\cal T}_{-}^{*}=1$     
         & \qquad & $\Lambda^{8} + \Lambda^{11}$
        & \qquad & \makecell{2$\times$3
        \\
        $(\nu, \eta, \kappa)$
        }
        & \qquad & 2$\times$15~
        \makecell{
        $(
d_{1Ra},d_{1Rs},
d_{2Ra},d_{3Ra},d_{3Rs},
d_{4Ra},d_{4Rs},d_{5Ra},
d_{6Ra},d_{6Rs},
 $\\$
d_{7Ra},d_{7Rs},
d_{8Ra},d_{9Ra},d_{9Rs},
d_{10Ra},d_{10Rs},
d_{11Ra},d_{12Ra},d_{12Rs},
 $\\$
d_{13Ra},d_{13Rs},d_{14Ra},d_{14Rs},
d_{15Ra},d_{15Rs},
d_{1Ia},d_{1Is},
d_{2Is}, d_{3Ia},d_{3Is},
 $\\$
d_{4Ia},d_{4Is},d_{5Is},
d_{6Ia},d_{6Is},d_{7Ia},d_{7Is},
d_{8Is}, d_{9Ia},d_{9Is},
d_{10Ia},d_{10Is},
$\\$
d_{11Is},
d_{12Ia},d_{12Is},d_{13Ia},d_{13Is},
d_{14Ia},d_{14Is},d_{15Is}, d_{15Ia}        
        )$
        }
        \\
         \hline 
       PHS$^{\dagger}$ with ${\cal T}_{-}{\cal T}_{-}^{*}=-1$     
         & \qquad & $-\i (\Lambda^{1} + \Lambda^{6})$
        & \qquad & \makecell{2$\times$3
        \\
        $(\nu, \eta, \kappa)$
        }
        & \qquad & 2$\times$15~
        \makecell{$
d_{1Rs},
d_{2Ra},d_{2Rs},
d_{3Ra},d_{3Rs},d_{4Ra},d_{4Rs},
d_{5Ra},d_{5Rs},
,d_{6Rs}
d_{7Rs},
$\\$
d_{8Ra},d_{8Rs},
d_{9Ra},d_{9Rs},
d_{10Ra},d_{10Rs},
d_{11Ra},d_{11Rs},
d_{12Rs},
d_{13Rs},
$\\$
d_{14Rs},d_{14Ra}
,d_{15Ra}, d_{15Rs},
 d_{1Ia},
d_{2Ia},d_{2Is},
d_{3Ia},d_{3Is},
$\\$
d_{4Ia},d_{4Is},
d_{5Ia},d_{5Is}, d_{6Ia},
d_{7Ia},d_{8Ia},d_{8Is},
d_{9Ia},d_{9Is},
$\\$
d_{10Ia},d_{10Is},
d_{11Ia},d_{11Is}, d_{12Ia},
d_{13Ia},d_{14Ia},d_{14Is},d_{15Is},d_{15Ia}        
        $}
        \\
         \hline 
       TRS with ${\cal T}_{+}{\cal T}_{+}^{*}=1$    
         & \qquad &  $\Lambda^{8} + \Lambda^{11}$
        & \qquad & \makecell{2$\times$3
        \\
        $(\nu, \eta, \kappa)$
        }
        & \qquad & 2$\times$15~
        \makecell{
$
(
d_{1Ra},d_{1Rs}, d_{2Rs},
d_{3Ra},d_{3Rs},d_{4Ra},d_{4Rs},
d_{5Rs},
d_{6Ra},d_{6Rs},d_{7Ra},
   $\\$ 
d_{7Rs},
d_{8Rs},
d_{9Ra},d_{9Rs},d_{10Ra},d_{10Rs},
d_{11Rs}, d_{12Rs},d_{12Ra},
d_{13Ra},$\\$
d_{13Rs},
d_{14Ra},d_{14Rs},
d_{15Rs},d_{15Ra},
d_{1Is},d_{1Ia},
d_{2Ia}
d_{3Ia},d_{3Is},
$\\$
d_{4Ia},d_{4Is},
d_{5Ia},
d_{6Ia},d_{6Is},d_{7Ia},d_{7Is},
d_{8Ia},d_{9Ia},d_{9Is},
$\\$
d_{10Ia},d_{10Is}, d_{11Ia},
d_{12Ia},d_{12Is},d_{13Ia},d_{13Is},
d_{14Ia},d_{14Is},d_{15Ia}, d_{15Is}
)
$        
        }
         \\
          \hline 
         TRS with ${\cal T}_{+}{\cal T}_{+}^{*}=-1$    
         & \qquad & $-\i (\Lambda^{1} + \Lambda^{6})$
        & \qquad & \makecell{2$\times$3
        \\
        $(\nu, \eta, \kappa)$
        }
         & \qquad & 2$\times$15~
         \makecell{
$
(
d_{1Ra},
d_{2Ra},d_{2Rs},
d_{3Ra},d_{3Rs},
d_{4Ra},d_{4Rs},
d_{5Ra},d_{5Rs},
$\\$
d_{6Ra},d_{7Ra},
d_{8Ra},d_{8Rs},
d_{9Ra},d_{9Rs},
d_{10Ra}, d_{10Rs},
d_{11Ra},d_{11Rs}, 
d_{12Ra},$\\$
d_{13Ra},
d_{14Rs},d_{14Ra},
d_{15Ra},d_{15Rs},
d_{1Is},
d_{2Ia},d_{2Is},
d_{3Ia},d_{3Is},
$\\$
d_{4Ia},d_{4Is},
d_{5Ia},d_{5Is},
d_{6Is},
d_{7Is},
d_{8Ia},d_{8Is},
d_{9Ia},d_{9Is},
$\\$
d_{10Ia}, d_{10Is},
d_{11Ia},d_{11Is},
d_{12Is}d_{13Is},
d_{14Ia},d_{14Is},
d_{15Is},d_{15Ia}
)
$         
         }
         \\
          \hline 
       TRS$^{\dagger}$ with ${\cal C}_{+}{\cal C}_{+}^{*}=1$    
         & \qquad & $\Lambda^{8} + \Lambda^{11}$
        & \qquad & \makecell{2$\times$3
        \\
        $(\nu, \eta, \kappa)$
        }
        & \qquad & 2$\times$15~
        \makecell{
$
(
d_{1Ra},d_{1Rs},
d_{2Rs},
d_{3Ra},d_{3Rs},
d_{4Ra},d_{4Rs},
d_{5Rs},
d_{6Ra},d_{6Rs},
$\\$
d_{7Ra},d_{7Rs},
d_{8Rs},
d_{9Ra},d_{9Rs},
d_{10Ra},d_{10Rs},
d_{11Rs},
d_{12Ra},d_{12Rs},
$\\$
d_{13Ra},d_{13Rs},
d_{14Ra},d_{14Rs},
d_{15Rs},d_{15Ra},
d_{1Ia},d_{1Is},
d_{2Is},
d_{3Ia},d_{3Is},
$\\$
d_{4Ia},d_{4Is},
d_{6Ia},d_{6Is},
d_{5Is},
d_{7Ia},d_{7Is},
d_{8Is},
d_{9Ia},d_{9Is},
$\\$
d_{10Is},d_{10Ia},
d_{11Is},
d_{12Ia},d_{12Is},
d_{13Ia},d_{13Is},
d_{14Ia},d_{14Is},
d_{15Is},d_{15Ia}
)
$        
        }
        \\
         \hline 
       TRS$^{\dagger}$ with ${\cal C}_{+}{\cal C}_{+}^{*}=-1$    
         & \qquad & $-\i (\Lambda^{1} + \Lambda^{6})$
        & \qquad & \makecell{2$\times$3
        \\
        $(\nu, \eta, \kappa)$
        }
        & \qquad & 2$\times$15~
        \makecell{
        $
d_{1Ra},
d_{2Ra},d_{2Rs},
d_{3Ra},d_{3Rs},
d_{4Ra},d_{4Rs},
d_{5Ra},d_{5Rs},
$\\$
d_{6Ra},
d_{7Ra},
d_{8Ra},d_{8Rs},
d_{9Ra},d_{9Rs},
d_{10Ra},d_{10Rs},
$\\$
d_{11Ra},d_{11Rs},
d_{12Ra},
d_{13Ra},
d_{14Rs},d_{14Ra},
d_{15Ra},d_{15Rs},
$\\$
d_{1Ia},
d_{2Ia},d_{2Is},
d_{3Ia},d_{3Is},
d_{4Ia},d_{4Is},
d_{5Ia},d_{5Is},
$\\$
d_{6Ia},
d_{7Ia},
d_{8Ia},d_{8Is},
d_{9Ia},d_{9Is},
d_{10Ia},d_{10Is},
$\\$
d_{11Ia},d_{11Is},
d_{12Ia},
d_{13Ia},
d_{14Is},d_{14Ia},
d_{15Ia},d_{15Is}        
        $
        }    
         \\
          \hline 
     \end{tabular}
     \vspace{1ex}

     {\raggedright Here $d_{\cal O}= d_{{\cal O}R} + \i d_{{\cal O}I}$ with ${\cal O} \in \{ x,y,z\}$. Symmetric and antisymmetric components of $d_{\cal O}$ with respect to $\bolds{k} \to -\bolds{k}$ are labelled by $d_{{\cal O}s}$ and $d_{{\cal O}s}$, respectively. $\eta, \nu,$ and $\kappa$ are introduced in Eq.~(\ref{eq:eta-4band}, \ref{eq:nu-4band}, \ref{eq:kappa-4band}). Note that nonzero parameters might vary by changing the depicted generators for each symmetry operator. Nevertheless, the number of parameters and constraints remain intact.   \par}
\end{table*}

\begin{table*}
     \centering
    \caption{\label{tab:sympar4II}Number of constraints and parameters to realize degenerate points in 4-band systems}
     \begin{tabular}{l|ll|ll|ll}
        \hline \hline 
        Symmetry  &\quad & Operator &\quad & $\#$ constr.  &\quad  & $\#$ parameters \\
        \hline 
        \hline
        CS
         & \qquad & $\Gamma_{5}$
        & \qquad & \makecell{3\\~
        $(\nu_{R}, \eta_{R},\kappa_{I} )$}
        & \qquad & 26
        \makecell{
$
(
d_{1R},
d_{3R},
d_{4R},
d_{6R},
d_{7R},
d_{8R},
d_{9R},
d_{10R},
$\\$
d_{11R},
d_{12R},
d_{13R},
d_{14R},
d_{15R},
d_{1I},
d_{2I},
d_{3I},
$\\$
d_{4I},
d_{5I},
d_{6I},
d_{7I},
d_{9I},
d_{10I},
d_{12I},
d_{13I},
d_{14I},d_{15I}
)
$        
        }
        \\
        \hline
                 ${\rm ps  CS}$ 
           & \qquad & $\Gamma_{5}$
           & \qquad & \makecell{4\\$(
           \eta,\nu)$}
           & \qquad &  $2 \times 15$\makecell{
           $(
           d_{1R},
           d_{2R}, 
          d_{3R}, 
           d_{4R},
           d_{5R}, 
           d_{6R},
           d_{7R}, 
           d_{8R},
           d_{9R},
           d_{10R},
           $\\$
           d_{11R},
           d_{12R},
           d_{13R},  
           d_{14R},    
           d_{15R},   
d_{1I},
d_{2I},
d_{3I},
d_{4I},
d_{5I},
d_{6I},
$\\$
 d_{7I},
d_{8I},
 d_{9I}, 
 d_{10I},
 d_{11I},
 d_{12I}, 
  d_{13I},   
   d_{14I},
   d_{15I}
           )$}
          \\
         \hline 
         SLS
         & \qquad & $\i \Gamma_{5}$
        & \qquad & \makecell{4\\$(
           \eta,\nu)$}
        & \qquad & 26~
        \makecell{
$(
d_{1R},
d_{3R},
d_{4R},
d_{6R},
d_{7R},
d_{8R},
d_{9R},
d_{10R},
$\\$
d_{11R},
d_{12R},
d_{13R},
d_{14R},
d_{15R},
d_{1I},
d_{3I},
d_{4I},
d_{6I},
$\\$
d_{7I},
d_{8I},
d_{9I},
d_{10I},
d_{11I},
d_{12I},
d_{13I},
d_{14I},
d_{15I}
)$        
        }
        \\
         \hline 
       $\cal I$ symmetry
         & \qquad & $ \Lambda^{13} -\frac{\Lambda^{14}}{\sqrt{3}} + \sqrt{\frac{2}{3}} \Lambda^{15} $
        & \qquad &  \makecell{2$\times$3
        \\
        $(\nu, \eta, \kappa)$
        }
        & \qquad & 2$\times$15~
        \makecell{
        $(
d_{1Ra},
d_{2Rs},
d_{3Ra},
d_{4Ra},
d_{5Rs},
d_{6Ra}
d_{7Ra},
d_{8Rs},
d_{9Ra},
d_{10Ra},
$\\$
d_{11Rs},
d_{12Ra},
d_{13Rs},
d_{14Rs},
d_{15Rs},
d_{1Is},
d_{2Ia},
d_{3Is},
d_{4Is},
d_{5Ia},
$\\$
d_{6Is},
d_{7Is},
d_{8Ia},
d_{9Is},
d_{10Is},
d_{11Ia},
d_{12Is},
d_{13Ia},
d_{14Ia},
d_{15Ia}        
        )$
        }
        \\
         \hline 
       $\rm psH$
         & \qquad &  $\Gamma_{1}$
        & \qquad &  \makecell{3 \\($\eta_{R}, \nu_{R} ,\kappa_{R}$)}

        & \qquad & 26~  
        \makecell{
$
(
d_{1R},
d_{3R},
d_{4R},
d_{6R},
d_{7R},
d_{8R},
d_{9R},
d_{10R},
$\\$
d_{11R},
d_{12R},
d_{13R},
d_{14R},
d_{15R},
d_{1I},
d_{2I},
d_{3I},
$\\$
d_{4I},
d_{5I},
d_{6I},
d_{7I},
d_{9I},
d_{10I},
d_{12I},
d_{13I},
d_{14I},
d_{15I}
)
$        
        }
        \\
         \hline 
        $\cal P$ symmetry
         & \qquad & $ \Lambda^{13} -\frac{\Lambda^{14}}{\sqrt{3}} + \sqrt{\frac{2}{3}} \Lambda^{15} $
        & \qquad & \makecell{2$\times$3
        \\
        $(\nu, \eta, \kappa)$
        }
        & \qquad & 2$\times$15~\makecell{$
        (
d_{1Ra},
d_{2Rs},
d_{3Ra},
d_{4Ra},
d_{5Rs},
d_{6Ra},
d_{8Rs},
d_{9Ra},
d_{7Ra},
d_{10Ra},
$\\$
d_{11Rs},
d_{12Ra},
d_{13Rs},
d_{14Rs},
d_{15Rs},
d_{1Ia},
d_{2Is},
d_{3Ia},
d_{4Ia},
d_{5Is},
$\\$
d_{6Ia},
d_{7Ia},
d_{8Is},
d_{9Ia},
d_{10Ia},
d_{11Is},
d_{12Ia},
d_{13Is},
d_{14Is},
d_{15Is} 
)
        $}
        \\
         \hline 
         $\cal PT$ symmetry
         & \qquad & ${\cal P} \times (\Lambda^{8}+\Lambda^{11})$
        & \qquad &  \makecell{3 \\
        $(\eta_{R}, \nu_{R}, \kappa_{R})$}
        & \qquad & 26~
        \makecell{
$(
d_{1R},
d_{2R},
d_{3R},
d_{4R},
d_{5R},
d_{6R},
d_{7R},
d_{8R},
d_{9R},
$\\$
d_{10R},
d_{11R},
d_{12R},
d_{13R},
d_{14R},
d_{15R},
d_{3I},
d_{4I},
$\\$
d_{1I},
d_{6I},
d_{7I},
d_{9I},
d_{10I},
d_{12I},
d_{13I},
d_{14I},
d_{15I}
)
$        
        }
\\
        \hline 
         ${\cal  CP}$ symmetry 
           & \qquad & $(\Lambda^{8}-\Lambda^{11})$
           & \qquad & \makecell{3\\$(\eta_{R}, \nu_{R}, \kappa_{I})$}
           & \qquad & 26~
           \makecell{$
           (
           d_{1R}, 
           d_{3R},
           d_{4R},
           d_{6R},  
           d_{7R},  
           d_{9R}, 
           d_{10R},
           $\\$
           d_{12R},
           d_{13R},
           d_{14R},
           d_{15R}, 
           d_{1I},
           d_{2I}, 
           d_{3I},
           d_{4I},
           d_{5I},
           $\\$
           d_{6I},
           d_{7I},
           d_{8I},
           d_{9I},
           d_{10I},
            d_{11I},
           d_{12I},
           d_{13I}, 
            d_{14I},
             d_{15I})
           $}
          \\
        \hline 
     \end{tabular}
     \vspace{1ex}

     {\raggedright Here $d_{\cal O}= d_{{\cal O}R} + \i d_{{\cal O}I}$ with ${\cal O} \in \{ x,y,z\}$. Symmetric and antisymmetric components of $d_{\cal O}$ with respect to $\bolds{k} \to -\bolds{k}$ are labelled by $d_{{\cal O}s}$ and $d_{{\cal O}s}$, respectively. $\Gamma_{1}$ and $\Gamma_{5}$ are given in Eqs~(\ref{eq:G1}, \ref{eq:G5}). $\eta, \nu$ and $\kappa$ are introduced in Eq.~(\ref{eq:eta-4band}, \ref{eq:nu-4band}, \ref{eq:kappa-4band}). Note that nonzero parameters might vary by changing the depicted generators for each symmetry operator. Nevertheless, the number of parameters and constraints remain intact. \par}
\end{table*}

We now turn to EP4s and present the most generic four-band Hamiltonian decomposed in the generalized Gell-Mann basis
\begin{align}
{\cal H}(k) =  d_{0}(k) \id_{4} +\bolds{d}(k) \cdot \bolds\Lambda ,
\label{eq:H4d}
\end{align}
where $\bolds\Lambda=(\Lambda_1, \Lambda_2, \ldots, \Lambda_{15})$ is the vector of four-band Gell-Mann matrices (see Appendix~\ref{app:gellmann_su4}), $\id_{4}$ is the $4 \times 4$ identity matrix, $\bolds{k}$ denotes the momentum with the appropriate dimensions, and $( d_{0}(\bolds{k}) ,\bolds{d}(\bolds{k}) )$ are complex-valued momentum dependent variables. 

The associated characteristic polynomial for $\cal H$ in Eq.~\eqref{eq:H4d}, from Eq.~\eqref{eq:charac_poly_n}, is given by
\begin{align}
{\cal F}_{\lambda}&= \lambda^{4} - a \lambda^{3} + b \lambda^{2} 
- c \lambda +d=0, \label{eq:char_poly_4}
\end{align}
where
\begin{align}
    a&=\tr[{\cal H}] , \label{eq:4band-a}\\
    b&=\frac{(\tr[{\cal H}])^{2} - \tr[{\cal H}^2]}{2} ,\label{eq:4band-b}\\
    c&= \frac{ \left(\tr[{\cal H}]^3-3 \tr[{\cal H}] \tr[{\cal H}^{2}]+2 \tr[{\cal H}^{3}]\right) }{6} ,\label{eq:4band-c}\\
    d &= \det[{\cal H}]. \label{eq:4band-d}
\end{align}
The four solutions $\lambda_1, \lambda_2, \lambda_3, \lambda_4$ of ${\cal F}_{\lambda}$ are eigenvalues of $\cal H$ in Eq.~\eqref{eq:H4d} and are given explicitly in Appendix~\ref{app:eigs_four_band}.
The discriminant associated with Eq.~\eqref{eq:char_poly_4} reads
\begin{equation}
    \mathcal{D} =
\frac{4 \eta^{3} - \nu^{2}}{27},
\end{equation}
where $\eta$, $\nu$ and $\kappa$ are
\begin{align}
\eta &= -3 a c+b^2+12 d ,  \label{eq:eta-4band}\\
\nu &= 27 a^2 d-9 a b c+2 b^3-72 b d+27 c^2 , \label{eq:nu-4band} \\
\kappa &= a^3-4 a b+8 c, \label{eq:kappa-4band}
\end{align}
with $a,b,c,d$ in Eqs.~(\ref{eq:4band-a},\ref{eq:4band-b},\ref{eq:4band-c},\ref{eq:4band-d}), respectively.
From the structure of this discriminant, one may naively expect that merely \emph{four} real constraints, namely $\Re[\eta]=\Im[\eta]=\Re[\nu]=\Im[\nu]=0$, should be satisfied to observe EP4s in 4-band systems. However, to force all roots of $\mathcal{D}$ to coincide, a third constraint, namely $\kappa$ in Eq.~\eqref{eq:kappa-4band}, should also be set to zero. This can be better understood if we follow the argument mentioned in Sec.~\ref{sec:nEps} by counting numbers of available traces~($\tr[{\cal H}^{2}], \tr[{\cal H}^{3}]$) and the determinant~($\det[{\cal H}]$) in the companion matrix of EP4s given by
\begin{align}
    {\cal H}_{0}=
    \begin{pmatrix}
    0 & 1 & 0 & 0 \\
    0 & 0 & 1 & 0 \\
    0 & 0 & 0 & 1 \\
    -\det[{\cal H}] & \frac{\tr[{\cal H}^{3}]}{3} & \frac{ \tr[{\cal H}^2]}{2} & 0
    \end{pmatrix}.
    \label{eq:H0-4band}
\end{align}
 Note that, without loss of generality, we set $\tr[{\cal H}] = 0$ as before. As a result, \emph{six} real constraints should be imposed to obtain EP4s in a four-band system. We summarize these constraints in the presence of various symmetries in Tables~\ref{tab:sympar4I} and \ref{tab:sympar4II}. Here aside from considering $\Lambda$ matrices as symmetry generators, we also use two Gamma matrices (cf. Appendix~\ref{app:gellmann_su4}) defined as
\begin{align}
    \Gamma_{1} &= \sigma_{x} \otimes \id_{2 },\label{eq:G1}\\
    \Gamma_{5} &= \sigma_{y} \otimes \tau_{z},\label{eq:G5}
\end{align}
where $\bolds{\sigma}$ and $\bolds{\tau}$ are Pauli matrices. We again note that in the case of a Hermitian model ($\bolds{d}_I = 0$), a four-band crossing requires solving 15 constraints $\bolds{d}_R = 0$.

Perturbing in the vicinity of EP4s with $\tr[{\cal H}]=0$ is described by ${\cal H}_{0}$ in Eq.~\eqref{eq:H0-4band}. To find various types of EP4s, we consider different cases, summarized in Table~\ref{tab:Ep4s}:
i) For Hamiltonians with $\tr[{\cal H}^2]=\tr[{\cal H}^3]=0$ the energy dispersion close to EP4s casts $ \i^{k} \sqrt[4]{ \det[{\cal H}]}$  with $k=1,2,3,4$.
ii) If the Hamiltonian is constructed in such a way that $\det[{\cal H}]=\tr[{\cal H}^2]=0$, the energy dispersion of ${\cal H}_{0}$ reads $0, (-1)^{j+j/3} \sqrt[3]{\tr[{\cal H}^{3}]/3}$ with $j=1,2,3$.
iii) When $\det[{\cal H}]=\tr[{\cal H}^3]=0$ for a 4-band Hamiltonian, the system exhibits two flat bands with energy zero and two dispersive bands $ \pm \sqrt{ \tr[{\cal H}^2]/2}$.
iv) The fourth situation is when close to an EP4, $\eta$ and $\kappa$ given in Eqs.~(\ref{eq:eta-4band}, \ref{eq:kappa-4band}) decrease to zero faster that $\nu$ in Eq.~\eqref{eq:nu-4band}. In this case, the low-energy dispersion should be computed from the general eigenvalues given in Eqs.~(\ref{eq:lam1-lvl4}, \ref{eq:lam2-lvl4}, \ref{eq:lam3-lvl4}, \ref{eq:lam4-lvl4}). As a result, the four bands close to the EP4 are proportional to $\mp \sqrt{2} \sqrt{-8b - 2^{2/3} \sqrt[3]{\nu}} \mp \sqrt{-8b + 2^{5/3} \sqrt[3]{\nu}}$.

\begin{table}
     \centering
    \caption{\label{tab:Ep4s}Various possibilities of EP4s and their energy dispersion}
     \begin{tabular}{l|ll|ll}
     \hline \hline 
           & \quad& Condition  & \quad& Energy dispersion \\
        \hline 
        \hline
        EP4 0 \quad&  & $\eta\neq 0, \nu\neq 0, \kappa \neq 0$ &\quad&  $ (\lambda_{1}, \lambda_{2}, \lambda_{3}, \lambda_{4})$ 
        \\
        & \quad& &\quad&\\
        EP4 I \quad&  & $\tr[{\cal H}^2]=\tr[{\cal H}^3]=0$ &\quad&  $ \i^{k} \sqrt[4]{ \det[{\cal H}]}$ 
        \\
        & \quad& &\quad&\\
        EP4 II \quad&  & $\det[{\cal H}]=\tr[{\cal H}^2]=0$ & \quad&  $ 0, (-1)^{j+j/3} \sqrt[3]{\tr[{\cal H}^{3}]/3}$
        \\
          & \quad& &\quad&\\
        EP4 III \quad&  & $\det[{\cal H}]=\tr[{\cal H}^3]=0$ & \quad&  $ 0,0, \pm \sqrt{ \tr[{\cal H}^2]/2}$ \\
          & \quad& &\quad&\\
        EP4 IV \quad&  & $\eta, \kappa\to 0$ faster than $\nu \to 0$ & \quad&  \makecell{$\pm \sqrt{2} \omega_{1} \pm \omega_{2}$
        } 
        \\
         \hline 
     \end{tabular}
      {\raggedright Here $k \in \{1,2,3,4\}$, $j \in \{1,2,3\}$, $\omega_{1}=\sqrt{-8b - 2^{2/3} \sqrt[3]{\nu}}$ and $\omega_{2} = \sqrt{-8b + 2^{5/3} \sqrt[3]{\nu}}$ with $b$ given in Eq.~\eqref{eq:4band-b}. Note that in all cases \emph{six} real constraints should be satisfied to observe EP4s. These constraints are counted by three complex equations either  $(\eta=0,\nu=0, \kappa=0)$ or $(\tr[{\cal H}^{2}]=0, \tr[{\cal H}^{3}]=0, \det[{\cal H}]=0)$. $ (\lambda_{1}, \lambda_{2}, \lambda_{3}, \lambda_{4})$ are given in Eqs.~(\ref{eq:lam1-lvl4}, \ref{eq:lam2-lvl4}, \ref{eq:lam3-lvl4}, \ref{eq:lam4-lvl4})  
      .\par}
\end{table}

Aside from EP4s, one might also encounter EP3s and EP2s in four-band systems. Let us first consider the case in which EP3s can be realized. The effective Hamiltonian reads
\begin{align}
    {\cal H}_{1} =
    \begin{pmatrix}
    a & 0 & 0 & 0 \\
    0 & b & c & d \\
    0 & e & f & g \\
    0 & h & i & j
    \end{pmatrix}=
    \begin{pmatrix}
    a & 0_{1 \times 3} \\
    0_{3 \times 1} & h_{3 \times 3}
    \end{pmatrix}.
\end{align}
Without loss of generality we consider $\tr[h_{3 \times 3}]=b+f+j=0$. Based on our results in Sec.~\ref{sec:3lvl}, we conclude that $h_{3 \times 3}$ can host EP3s if $\eta$ and $\nu$ given in Eqs.~(\ref{eq:eta3lvl}, \ref{eq:nu3lvl}) with ${\cal H}=h_{3 \times 3}$ are simultaneously zero.

To explore EP2s in four-band systems, we consider two possibilities: a four-band system with i) two trivial bands and an EP2, or ii) two EP2s.
For the former scenario, we introduce a generic Hamiltonian which reads
\begin{align}
    {\cal H}_{2} &=
    \begin{pmatrix}
         a & 0 & 0 & 0 \\
         0 & b & 0 & 0 \\
         0 & 0 & c & d \\
         0 & 0 & e & f \\
    \end{pmatrix}=
    \begin{pmatrix}
    a & 0 & 0_{1 \times 2} \\
    0 & b & 0_{1 \times 2} \\
    0_{2 \times 1} & 0_{2 \times 1} & h_{2 \times 2}\\
    \end{pmatrix}.
\end{align}
Following the results in Sec.~\ref{sec:2lvl}, we conclude that ${\cal H}$ possesses an EP2 when $\eta$ and $\nu$ constraints in Eq.~\eqref{eq:cond2b} are satisfied by $h_{2 \times 2}$.
The second plausible situation to detect EP2s can be described by an effective Hamiltonian given by
\begin{align} \label{eq:ep2s_in_4_band_model}
{\cal H}_{3} &=
    \begin{pmatrix}
    a & b & 0 & 0 \\
    c & d & 0 & 0 \\
    0 & 0 & e & f \\
    0 & 0 & g & h
        \end{pmatrix}=
           \begin{pmatrix}
    \tilde{h}_{2 \times 2} & 0_{2 \times 2} \\
    0_{2 \times 2} &  \overline{h}_{2 \times 2}\\
    \end{pmatrix}. 
\end{align}
${\cal H}_{3}$ displays EP2s if discriminants of $ \tilde{h}_{2 \times 2}$ and $\overline{h}_{2 \times 2}$ are set to zero, i.e., Eq.~\eqref{eq:cond2b} is satisfied for $\tilde{h}_{2 \times 2}$ and $\overline{h}_{2 \times 2}$. In very special cases in which both discriminants acquire zero in a particular parameter regime, we can realize the coexistence of two EP2s.

\begin{figure}
    \centering
    \includegraphics[width=0.9\columnwidth]{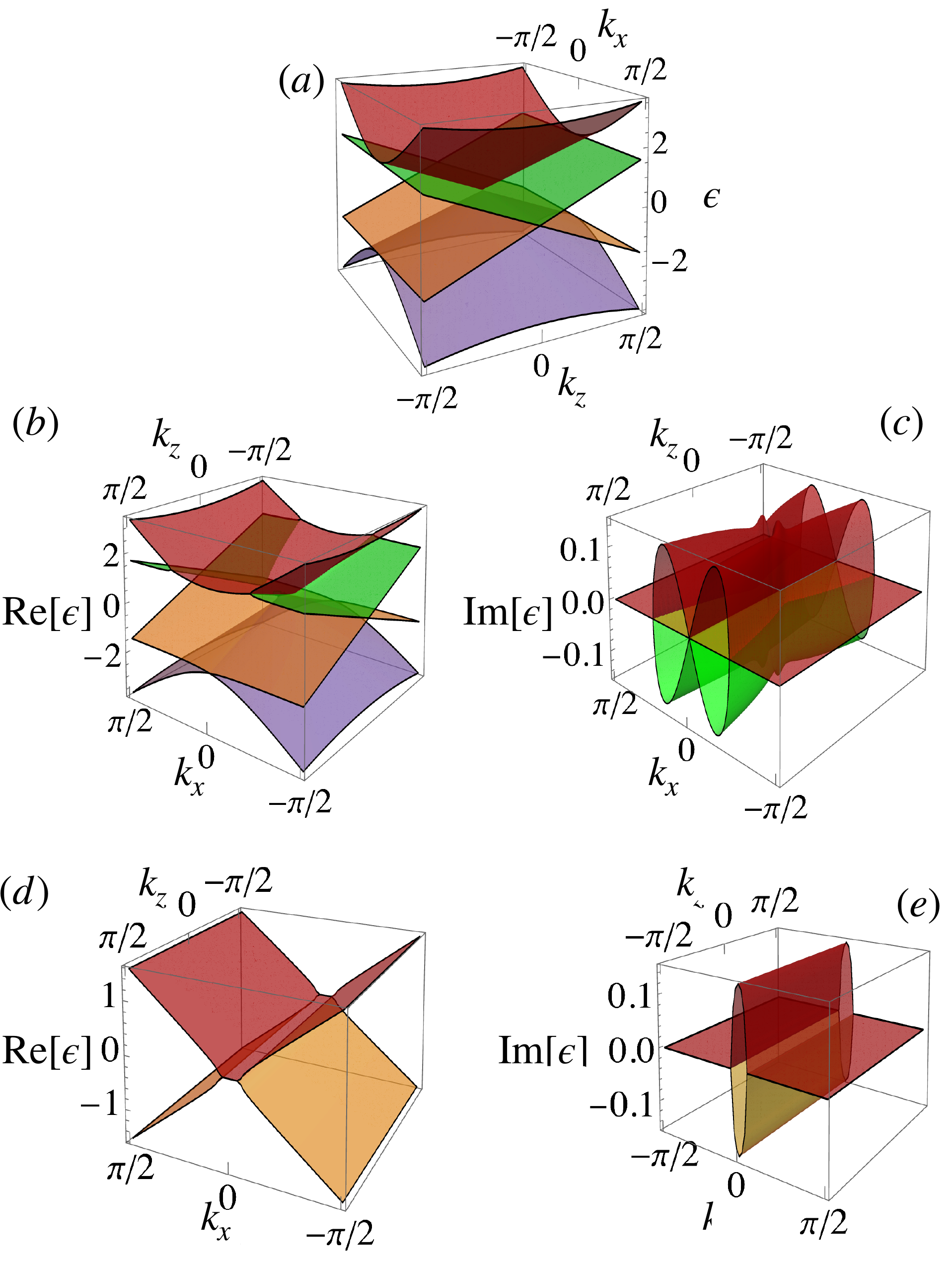}
    \caption{(a) The spectrum of the four-band model in Eq.~\eqref{eq:Hex4band1} in its Hermitian limit with $\alpha_{p}=\alpha_{m}=\alpha_{z}=\alpha_{b}=0$ and $\theta_{1}=\theta_{2}=\pi/2$. (Middle panels) The real~(b) and imaginary~(c) components of the band structure for the non-Hermitian model in Eq.~\eqref{eq:Hex4band1} with $\alpha_{p}=\alpha_{m}=0.15$, $\alpha_{z}= 0.15\i$, and $\alpha_{b}=0$.
    (Bottom panels) The real~(d) and imaginary~(e) components of the band structure for the non-Hermitian model in the presence of psH symmetry in Eq.~\eqref{eq:Hex4band_psh}. Bands in panels (d-e) are twofold degenerate. Line colors in middle and bottom panels are chosen such that lowest to higher bands are presented in blue, orange, green, and red colors, respectively.
    }
    \label{Fig:Hex14band}
\end{figure}

\begin{figure}
    \centering
    \includegraphics[width=0.85\columnwidth]{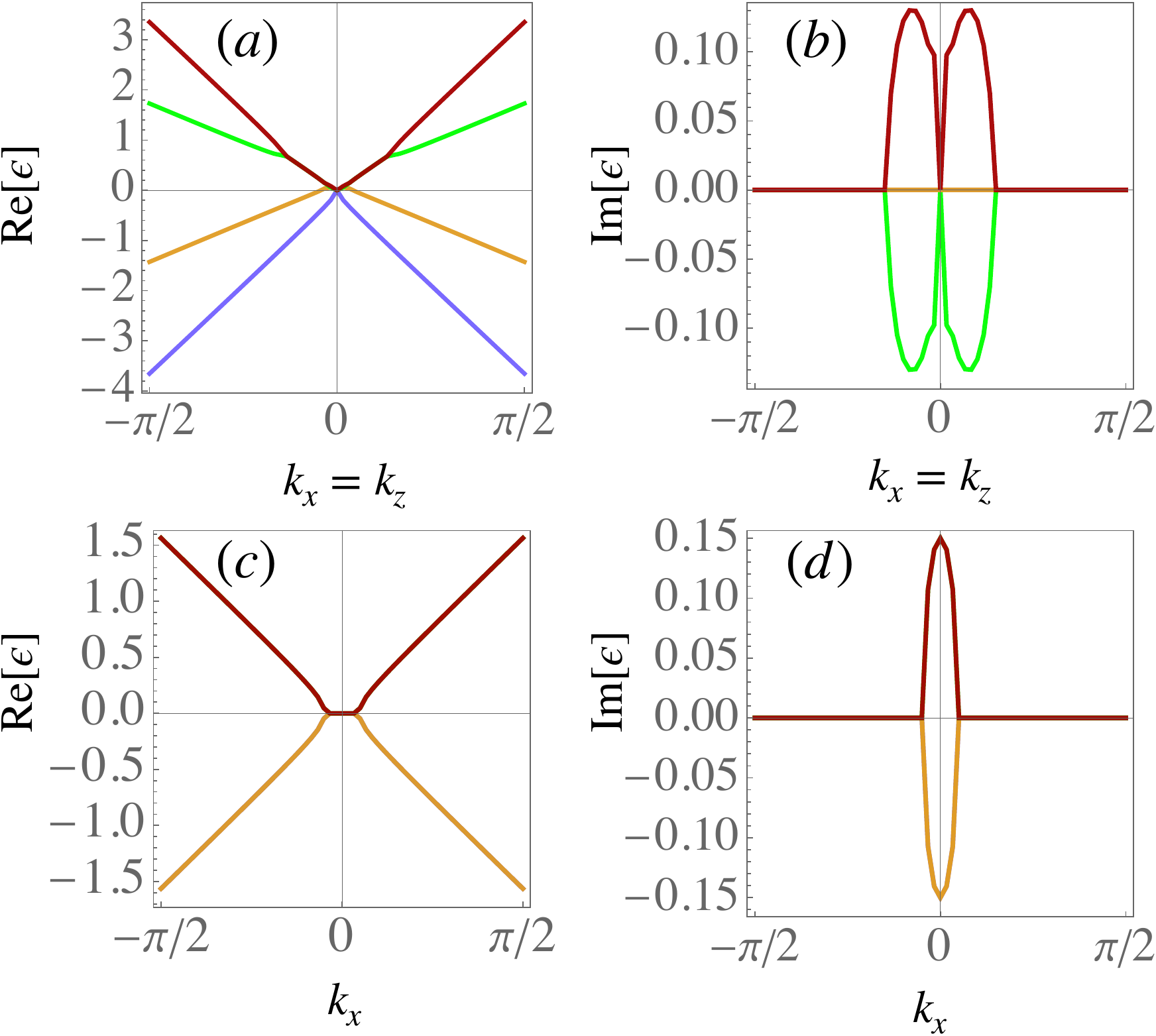}
    \caption{ (Upper panels) The same as panels (b,c) in Fig.~\ref{Fig:Hex14band} along the $k_{x}=k_{z}$ direction.
    (Bottom panels) The same as panels (d,e) in Fig.~\ref{Fig:Hex14band} at $k_{z}=0$ and along the $k_{x}$ direction.
    }
    \label{Fig:Hex14band_cut}
\end{figure}

To exemplify the role of symmetries on the low-energy dispersion of EP4s, we present a case study in the following.
We start with a traceless non-Hermitian four-band model, which reads 
\begin{align}
    {\cal H} 
    =
\left(
\begin{array}{cccc}
 0 
 &
 \alpha_{p}+ k_{x} 
 &
h_{zz2}
 &
h_{bx}
\\
  k_{x}-\alpha_{p} 
 &
 0 
 &
\tilde{h}_{bx2}&
h_{zz1}
 \\
\tilde{h}_{zz2}
 &
h_{bx2}
 &
 0 
 &
  k_{x}-\alpha_{m} 
 \\
\tilde{h}_{bx}
&
\tilde{h}_{zz1} &
 \alpha_{m}+ k_{x} 
 &
 0 \\
\end{array}
\right)
,\label{eq:Hex4band1}
\end{align}
where $h_{zz2}= \alpha_{z}- e^{\i \theta_{1}} k_{z}  $, $\tilde{h}_{zz2}= -\alpha_{z}- e^{-\i \theta_{1}} k_{z} $, $h_{bx}= \alpha_{b}+ e^{-\i \theta_{2}} k_{x} $, $\tilde{h}_{bx}= -\alpha_{b}+ e^{\i \theta_{2}} k_{x} $, $\tilde{h}_{bx2}= -\alpha_{b}+ e^{-\i \theta_{2}} k_{x} $, $h_{bx2}= \alpha_{b}+ e^{\i \theta_{2}} k_{x} $, $h_{zz1}= \alpha_{z}+ e^{\i \theta_{1}} k_{z} $, and $\tilde{h}_{zz1}= -\alpha_{z}+ e^{-\i \theta_{1}} k_{z}$. Here $\alpha_{\cal O}$ with ${\cal O} \in \{ p,m,z,b \}$ are complex-valued non-Hermitian parameters, $(\theta_{1}, \theta_{2})$ denote phase variables, and $(k_{x}, k_{z})$ indicate momenta. When non-Hermitian variables $\alpha_{\cal O}$ vanish, the Hamiltonian in Eq.~\eqref{eq:Hex4band1} describes the low-energy band structure of four-fold fermions at $k_{y}=0$~\cite{Jin2021}. We plot the dispersion relation for this four-band model in the Hermitian limit with $\alpha_{p}=\alpha_{m}=\alpha_{z}=\alpha_{b}=0$ and $\theta_{1}=\theta_{2}=\pi/2$ in Fig.~\ref{Fig:Hex14band}(a). This Hermitian band structure displays a fourfold degeneracy in its spectrum at $k_{x}=k_{z}=0$. The traces and the determinant of the Hamiltonian at $\theta_{1}=\theta_{2}=\pi/2$ in Eq.~\eqref{eq:Hex4band1} read
\begin{align}
    \tr[{\cal H}] &= 0 ,\\
    \tr[{\cal H}^{2}] &= 
    -2 \left(2 \alpha_{b}^2+\alpha_{m}^2+\alpha_{p}^2+2 \alpha_{z}^2\right)+8 k_{x}^2+4 k_{z}^2
    ,\\
    \tr[{\cal H}^{3}] &=
    24 i \alpha_{z} k_{x}^2
    ,\\
    \det[{\cal H}] &=
    k_{x}^2 \left(-(\alpha_{m}-\alpha_{p})^2+4 \alpha_{z}^2+4 k_{z}^2\right)
    \nonumber \\
    &\quad
    +\left(\alpha_{b}^2+\alpha_{m} \alpha_{p}+\alpha_{z}^2+k_{z}^2\right)^2.
\end{align}
For simplicity purpose, we merely consider cases in which  $\alpha_{m}=\alpha_{p}=\alpha$, $\alpha_{z}=\i \alpha$, $\alpha_{b}=0$ with $\alpha$ be a real-valued number. In this parameter regime, constraints in Eqs.~(\ref{eq:eta-4band}, \ref{eq:nu-4band}, \ref{eq:kappa-4band}) cast
\begin{align}
\eta &=
\left(-4 k_{x}^2-2 k_{z}^2\right)^2+12 \left(k_{x}^2 \left(4 k_{z}^2-4 \alpha ^2\right)+k_{z}^4\right)
,\\
\frac{\nu}{2}&=
864 \alpha ^2 k_{x}^4+\left(-4 k_{x}^2-2 k_{z}^2\right)^3
\nonumber \\
&
-36 \left(-4 k_{x}^2-2 k_{z}^2\right) \left(k_{x}^2 \left(4 k_{z}^2-4 \alpha ^2\right)+k_{z}^4\right)
,\\
    \kappa &=
    -64 \alpha  k_{x}^2
    .
\end{align}
These constraints simultaneously vanish when $k_{x}=k_{z}=0$. As a result, EP4s appear in this system as we have shown in Figs.~\ref{Fig:Hex14band}(b,c) and Fig.~~\ref{Fig:Hex14band_cut}(a,b). As close to this EP4, $\eta$, $\nu$ and $\kappa$ are nonzero, we identify this EP4 as type 0, see Table~\eqref{tab:Ep4s}. Aside from EP4s, our model also exhibits EP2s close to $k_{x}=k_{z}\approx 0.47$, as shown in Figs.~\ref{Fig:Hex14band}(b,c) and Fig.~~\ref{Fig:Hex14band_cut}(a,b). 

To further explore the effect of symmetry on the appearance of EPs, we impose psH symmetry with generator $\Gamma_{1}$ on our Hamiltonian in Eq.~\eqref{eq:Hex4band1}. The psH-symmetric Hamiltonian then reads
\begin{align}
    {\cal H}_{\rm psH}=
    \left(
\begin{array}{cccc}
 0 
 &
 h_{1}
 &
h_{zz}
 &
 h_{x2} \\
h_{mpx}
 &
 0 
 &
 h_{x2}
 &
 \tilde{h}_{zz}
 \\
 -\tilde{h}_{zz}
 &
 h_{x2}
 &
 0 
 &
 h_{mpx}
 \\
 h_{x2}
  &
  -h_{zz}
  &
  h_{1} 
  &
  0 
  \\
\end{array}
\right),\label{eq:Hex4band_psh}
\end{align}
where $h_{1}= \frac{1}{2} (\alpha_{m}+\alpha_{p}+2  k_{x})$, $h_{zz}= \alpha_{z}- k_{z} \cos (\theta_{1}) $, $\tilde{h}_{zz}= \alpha_{z}+ k_{z} \cos (\theta_{1}) $ $h_{mpx}=-\frac{\alpha_{m}}{2}-\frac{\alpha_{p}}{2}+ k_{x} $, and $h_{x2}= k_{x} \cos (\theta_{2}) $. The associated characteristic polynomial at $\theta_{1}=\theta_{2}=\pi/2$ factorizes into two second-order polynomials as
\begin{align}
   \left(-\alpha ^2-\lambda ^2+k_{x}^2\right)^2 =0.
\end{align}
This twofold degeneracy is evident in Figs.~\ref{Fig:Hex14band}(d,e) and Figs.~\ref{Fig:Hex14band_cut}(c,d) in which we plot the band structure of ${\cal H}_{\rm psH}$ at $k_{z}=0$, $\alpha_{m}=\alpha_{p}=\alpha$, $\alpha_{z}=\i \alpha$, $\alpha_{b}=0$ with $\alpha=0.2$. Here we see that bands are doubly degenerate come in pairs as merely two bands are visible. The momenta at which EP2s occur are $k_{x}= \pm \alpha$. This can be obtained from the associated constraints for ${\cal H}_{\rm psH}$
\begin{align}
    \eta &= 16 \left(k_{x}^2-\alpha ^2\right)^2 ,\\
    \nu &= 128 \left(k_{x}^2-\alpha ^2\right)^3 ,\\
     \kappa &= 0 .
\end{align}
$\eta$ and $\nu$ are zero when $k_{x}= \pm \alpha$. Finally, in agreement with our findings, the number of constraints reduces when we impose psH symmetry to our non-Hermitian system in Eq.~\eqref{eq:Hex4band1}.

\section{Discussion and Conclusion}\label{sec:conclusion}

In this work, we have studied the appearance of exceptional points of any order in the presence of symmetries. In particular, we have addressed three questions pertaining to the number of constraints to find EP$n$s, the implications of symmetries on the number of constraints for realizing EPs, and the low-energy behavior of these EPs. By expressing the characteristic polynomial of an $n$-dimensional non-Hermitian Hamiltonian in terms of the determinant and traces of the Hamiltonian, we have shown that one can identify $2n-2$ real constraints for finding EP$n$s. We, furthermore, have discussed that in the presence of various symmetries, the number of constraints may reduce. Our results show that combining symmetries generally results in further decreasing the number of constraints.
By interpreting the companion matrix as a perturbation close to an EP$n$, we have explicitly identified plausible low-energy dispersions of EP$n$s.
Besides these general considerations for EPs of any order, we have derived exact results for EPs of orders two, three, and four.
Through looking at the companion matrix, we have also calculated explicit expressions for the dispersion around an EP, allowing us to characterize EP3s and EP4s based on their low-energy spectrum. In addition, we have presented the appearance of lower-order EPs in $n$-dimensional models and find that EP2s can be realized in three-band systems, while both EP2s and EP3s can appear in four-band systems.

While we have focused on EP$n$s in this work, we emphasize that our results can be straightforwardly generalized to exceptional structures of higher dimensions. Associating a parameter with each constraint, we have shown that EP$n$s generally appear in $n-1$-dimensional setups in the presence of, e.g., psH symmetry. Consequently, exceptional one-dimensional lines or two-dimensional surfaces of order $n$ appear generically in $n$- and $n+1$-dimensional systems, respectively. In other words, the number of constraints is related to the \emph{codimension} of the exceptional structure, i.e., the difference between the total dimension of the system and the dimension of the exceptional structure, and our results can thus be readily applied to study the realization of higher-dimensional exceptional structures in the presence of symmetries.

Besides exceptional degeneracies, ordinary (Hermitian) degeneracies may appear in non-Hermitian systems where the eigenvalues coalesce, but the eigenbasis is complete. As we briefly discussed in Sects.~\ref{sec:2lvl}-\ref{sec:4lvl}, this requires setting ${\bf d}={\bf 0}$ for the various Hamiltonians such that these Hamiltonians are proportional to an identity matrix. Generally, this results in having to satisfy a large number of constraints to find these degeneracies. Indeed, one needs to satisfy $2 (n^2-1)$ constraints to find an ordinary $n$-fold degeneracy, where $n^2-1$ is the number of dimensions of the group SU($n$). Clearly, ${\bf d} = {\bf 0}$ is a solution to the characteristic polynomial in Eq.~\eqref{eq:charac_poly_n}. We note that one of the crucial differences between EPs and ordinary degeneracies on the level of polynomial equations sits in the relation between the characteristic and the minimal polynomials: For EPs, the characteristic polynomial equals the minimal polynomial, whereas, for ordinary degeneracies, the characteristic polynomial is a multiple of the minimal polynomial~\cite{lang02}.

In Ref.~\onlinecite{Delplace2021}, it is stated that symmetry-protected multifold exceptional points are points at which the symmetry is spontaneously broken. This is indeed the case for the symmetries the authors consider there (CS, psH, $\cal PT$, and $\cal CP$ symmetry), which are antiunitary symmetries that are local in parameter space. We here show that unitary, local symmetries such as SLS and psCS can also stabilize higher-order EPs in lower dimensions. These EPs do not mark a transition between broken and unbroken symmetry, thus showing that not all symmetry-protected EP$n$s necessarily correspond to spontaneous symmetry-breaking points.

While we have presented an extensive study here on the realization of exceptional points of any order in the presence of symmetry, we did not touch upon the possibility of defining topological invariants. Former studies proposed to define $\mathbb{Z}_2$ index based on either $\textrm{sign}(\det[\cal H])$ ($\textrm{sign}(\det[\i \cal H])$) in two-band models with $\cal PT$ ($\cal CP$) symmetry~\cite{Gong2018, Okugawa2019} or the sign of the discriminant~\cite{Delplace2021} for systems of any dimension with CS, psH, $\cal PT$, and $\cal CP$ symmetry. It would be intriguing to investigate whether more generic invariants could be defined based on our rigorous mathematical framework. We leave this problem for later studies.

\emph{Acknowledgments.---} We would like to thank Emil J. Bergholtz for pointing out Ref.~\onlinecite{Jiang2020}.

\bibliography{Higher_order_EP.bib}

\appendix
\newpage
\section{Non-Hermitian Bernard-LeClair symmetries} \label{app:symmBL}

Bernard and LeClair define non-Hermitian symmetries as follows~\cite{Bernard2002}.

\begin{itemize}
\item[i.] \emph{Q symmetry:}
\begin{align}
 {\cal H}(\bolds{k}) = \varepsilon_{q} q  {\cal H}^{\dagger}(\bolds{k}) q^{-1} , \quad q^{2}=1.
\end{align}
From the Q symmetry, we have
\begin{align}
 {\cal H}(\bolds{k}) q  | L_{n}(\bolds{k}) \rangle  &= \varepsilon_{q} q  {\cal H}^{\dagger}(\bolds{k})  | L_{n}(\bolds{k}) \rangle 
 ,\\
\Rightarrow \epsilon(\bolds{k}) &= \varepsilon_{q} \epsilon^{*}(\bolds{k}).
\end{align}
The discriminant of ${\cal H}$ given by ${\cal D}(k) := (-1)^{N(N-1)/2} \prod_{n \neq n'} (\epsilon_{n} -\epsilon_{n'})$ then mimics the behaviour of $\epsilon$ and reads
\begin{align}
{\cal D}(\bolds{k}) = \varepsilon_{q} {\cal D}^{*}(\bolds{k}).
\end{align}

\item[ii.] \emph{C symmetry:}
\begin{align}
{\cal H}(-\bolds{k}) = \varepsilon_{c} c {\cal H}^{T}(\bolds{k}) c^{-1} , \quad cc^{*} = \eta_{c} \id. 
\end{align}
where $\varepsilon_{c}, \eta_{c} \in \{ 1,-1\}$.
From the $\rm C$ symmetry, we have
\begin{align}
c^{-1} {\cal H}(-\bolds{k}) | R_{n}(\bolds{k}) \rangle  &= \varepsilon_{c}  {\cal H}^{T}(\bolds{k}) c^{-1} | R_{n}(\bolds{k}) \rangle 
\\
\Rightarrow 
\epsilon(-\bolds{k})  &= \varepsilon_{c} \epsilon(\bolds{k}).
\end{align}
To reach the last equality, we have used $(A-\lambda \id)^{T} =(A^{T} - \lambda \id) $ and $\det[(A-\lambda \id)^{T}]= \det[(A^{T} - \lambda \id)]$ which leads to the conclusion that $A, A^{T}$ have the same eigenvalues.
The discriminant of ${\cal H}$ then reads
\begin{align}
{\cal D}(-\bolds{k})  = \varepsilon_{c} {\cal D}(\bolds{k}).
\end{align}

\item[iii.] \emph{K symmetry:}
\begin{align}
{\cal H}(-\bolds{k}) = \varepsilon_{\kappa} \kappa {\cal H}^{*}(\bolds{k}) \kappa^{-1} , \quad \kappa \kappa^{*}=\eta_{\kappa} \id.
\end{align}
where $\varepsilon_{k}, \eta_{k} \in \{ 1,-1\}$.
From the $\rm K$ symmetry, we have
\begin{align}
 \kappa^{-1}   {\cal H}(-\bolds{k}) | R_{n}(\bolds{k}) \rangle  &= \varepsilon_{k}   {\cal H}^{*}(\bolds{k}) \kappa^{-1}  | R_{n}(\bolds{k}) \rangle 
 \\ \Rightarrow
 \epsilon(-\bolds{k}) &= \varepsilon_{k} \epsilon^{*}(\bolds{k}).
\end{align}
The discriminant of ${\cal H}$ then reads
\begin{align}
 {\cal D}(-\bolds{k}) = \varepsilon_{k} {\cal D}^{*}(\bolds{k})
.
\end{align}

\item[iv.] \emph{P symmetry:}
\begin{align}
 {\cal H}(\bolds{k}) = -p  {\cal H}(\bolds{k}) p^{-1} , \quad p^{2}=\id.
\end{align}
From the $\rm P$ symmetry, we have
\begin{align}
 {\cal H}(\bolds{k}) p  | R_{n}(\bolds{k}) \rangle  &= -p  {\cal H}(\bolds{k})   | R_{n}(\bolds{k}) \rangle 
\\
\Rightarrow
\epsilon(\bolds{k}) &= - \epsilon(\bolds{k}).
\end{align}
The discriminant of ${\cal H}$ then reads
\begin{align}
{\cal D}(\bolds{k}) = - {\cal D}(\bolds{k})
.
\end{align}

\end{itemize}

\begin{table}
     \centering
    \caption{\label{tab:BL}Summarized Bernard-LeClair symmetries and their associated energy constraints}
     \begin{tabular}{l|ll|ll}
        \hline \hline 
        Symmetry  &\qquad & Symmetry Constraint &\qquad  &Energy Constraint  \\
        \hline 
        \hline
          $\rm Q$ symmetry
        & \qquad & $ {\cal H}(\bolds{k}) = \varepsilon_{q} q  {\cal H}^{\dagger}(\bolds{k}) q^{-1} $
        & \qquad &  $\epsilon(\bolds{k}) = \varepsilon_{q} \epsilon^{*}(\bolds{k})$
        \\
          $\rm C$ symmetry
        & \qquad & $ {\cal H}(-\bolds{k}) = \varepsilon_{c} c  {\cal H}^{T}(\bolds{k}) c^{-1} $
        & \qquad & $\epsilon(-\bolds{k})  = \varepsilon_{c} \epsilon(\bolds{k})$
        \\
        $\rm K$ symmetry
        & \qquad & $v(-\bolds{k}) = \varepsilon_{k} \kappa  {\cal H}^{*}(\bolds{k}) \kappa^{-1} $
        & \qquad &  $ \epsilon(-\bolds{k}) = \varepsilon_{k} \epsilon^{*}(\bolds{k})$
        \\
         $\rm P$ symmetry
        & \qquad & $ {\cal H}(\bolds{k}) = -p  {\cal H}(\bolds{k}) p^{-1}  $
        & \qquad &  $\epsilon(\bolds{k}) = - \epsilon(\bolds{k})$
        \\
        \hline 
     \end{tabular}
     \vspace{1ex}

     {\raggedright Here $q^{2}=\id, cc^{*}=\eta_{c} \id, \kappa \kappa^{*}= \eta_{k} \id$, and $p^{2}=\id$. $\eta_{\cal O}, \varepsilon_{\cal O} \in  \{ 1,-1\}$.   \par}
\end{table}

The above four unitary matrices satisfy
\begin{align}
c= \varepsilon_{pc} p c p^{T}
,
\quad
\kappa=\varepsilon_{p\kappa} p  \kappa p^{T}
,
\\
c=\varepsilon_{qc} q c q^{T}
,
\quad 
p=\varepsilon_{pq} q p q^{\dagger},
\end{align}
where $\varepsilon_{pc}, \varepsilon_{p\kappa},\varepsilon_{qc},\varepsilon_{pq} 
\in \{-1,1 \}$~\cite{Liu2019}.

The energy constraints from this classification, summarized in Table~\ref{tab:BL}, is in agreement with our results from the other classification, summarized in Table~\ref{tab:symm}. Note that the nomenclature in these two classification are linked as follows. The $C$-symmetry corresponds to the $\rm PHS$/$\rm TRS^{\dagger}$, $Q$ symmetry corresponds to the $\rm CS$/$\rm psH$, $K$ symmetry is related to the $\rm TRS$/$\rm PHS^{\dagger}$, $P$ symmetry is the same as the $\rm SLS$. 

\section{Symmetry-allowed Hamiltonians} \label{app:symm_allo_hams}
We summarize Hamiltonians allowed by a specific symmetry, listed in Table~\ref{tab:symm}, in Table~\ref{tab:Hsymm}.

\begin{table*}
     \centering
    \caption{Summarized Hamiltonians with a particular symmetry}
     \begin{tabular}{l|ll}
        \hline \hline 
        Symmetry  &\qquad & Associated Hamiltonians   \\
        \hline 
        \hline
        Particle-hole symmetry I~(PHS)   
        &\qquad  & $
        {\cal H}_{\rm PHS} =\frac{1}{2}
        \Big[
        {\cal H}(-\bolds{k}) -
        {\cal C}_{-} {\cal H}^{T}(\bolds{k}) {\cal C}_{-}^{\dagger} 
       \Big]
        $ 
        \\
        \hline
        Particle-hole symmetry II~(PHS$^{\dagger}$)    
        & \qquad & $
        {\cal H}_{\rm PHS^{\dagger}} =\frac{1}{2}
        \Big[
        {\cal H}(-\bolds{k})
        -
        {\cal T}_{-} {\cal H}^{*}(\bolds{k}) {\cal  T}_{-}^{\dagger} 
        \Big]
        $       
        \\
        \hline
        Time-reversal symmetry I~(TRS)    
        & \qquad & $
        {\cal H}_{\rm TRS} =\frac{1}{2}
        \Big[
        {\cal H}(-\bolds{k})
        +
        {\cal T}_{+} {\cal H}^{*}(\bolds{k}) {\cal T}_{+}^{\dagger} 
        \Big]
        $        \\
        \hline
        Time-reversal symmetry II~(TRS$^{\dagger}$)    
        & \qquad & $
        {\cal H}_{\rm TRS^{\dagger}} =\frac{1}{2}
        \Big[
         {\cal H}(-\bolds{k})
         +
        {\cal C}_{+} {\cal H}^{T}(\bolds{k}) {\cal C}_{+}^{\dagger} 
        \Big]
        $       
 \\
 \hline
        Chiral symmetry~(CS)  
        & \qquad & $
        {\cal H}_{\rm CS} =\frac{1}{2}
        \Big[
        {\cal H}(\bolds{k})
        -
        \Gamma {\cal H}^{\dagger}(\bolds{k}) \Gamma^{-1} 
        \Big]
        $    

\\
\hline
        Pseudo-chiral symmetry~(psCS)  
        & \qquad & $
        {\cal H}_{\rm psCS} =\frac{1}{2}
        \Big[
        {\cal H}^T(\bolds{k})
        -
        \Lambda {\cal H}(\bolds{k}) \Lambda^{-1} 
        \Big]
        $    

\\
\hline
                 Sublattice-symmetry~(SLS)  
        & \qquad & $
        {\cal H}_{\rm SLS} =\frac{1}{2}
        \Big[
        {\cal S} {\cal H}(\bolds{k}) {\cal S}^{-1} -{\cal H}(\bolds{k})
        \Big]
        $  

 \\
 \hline
                Pseudo-Hermiticity~($\rm psH$) 
        & \qquad & $
        {\cal H}_{\rm psH} =\frac{1}{2}
        \Big[
        {\cal H}(\bolds{k})
        +
        \varsigma {\cal H}^{\dagger}(\bolds{k}) \varsigma^{-1} 
        \Big]
        $  

\\
\hline
               Inversion symmetry~($\cal I$)
        & \qquad & $
        {\cal H}_{\cal I} =\frac{1}{2}
        \Big[
        {\cal H}^{\dagger}(-\bolds{k})
        +
        {\cal I} {\cal H}(\bolds{k}) {\cal I}^{-1}  
        \Big]
        $ 

 \\
 \hline
         Parity~($\cal P$) symmetry
        & \qquad & $ {\cal H}_{\cal P} =\frac{1}{2}
        \Big[
        {\cal H}(-\bolds{k})
        +
        {\cal P} {\cal H}(\bolds{k}) {\cal P}^{-1}
        \Big]
        $    

 \\
 \hline
         Parity-time~($\cal PT$) symmetry
        & \qquad & $ {\cal H}_{\cal PT} =\frac{1}{2}
        \Big[
        {\cal H}(\bolds{k})
        +
        ({\cal P}{\cal T}_{+}) {\cal H}^{*}(\bolds{k}) ({\cal P}{\cal T}_{+})^{-1} 
        \Big]$   

\\
\hline
        Parity-particle-hole~(${\cal CP}$) symmetry
        & \qquad & ${\cal H}_{\cal CP} =\frac{1}{2}
        \Big[
        {\cal H}(\bolds{k})
        -
        ({\cal CP}) {\cal H}^{*}(\bolds{k}) ({\cal CP})^{-1} 
        \Big]$ 

\\
        \hline 
     \end{tabular}
     \vspace{1ex}

     {\raggedright Here the unitary operator $A \in \{ \Gamma, \Lambda, \varsigma, {\cal S}, {\cal P}, {\cal I} \}$ obeys $A^{2}=1$, and the anti-unitary operator $A \in \{ {\cal C}_{\pm}, {\cal T}_{\pm} \}$ satisfies $AA^{*}=\zeta_{A} 1$ with $\zeta_{A}= \pm 1$. \par}
     \label{tab:Hsymm}
\end{table*}

\section{General considerations for number of constraints to realize EP$n$s} \label{app:epn_symmetry}

In the main text, we present that $2(n-1)$ constraints should be satisfied to find an EP$n$. We further show that these constraints explicitly read $\Re[\det[{\cal H}]] = 0$, $\Im[\det[{\cal H}]] = 0$, $\Re[\tr[{\cal H}^2]] = 0$, $\Im[\tr[{\cal H}^2]] = 0$, $\ldots$, $\Re[\tr[{\cal H}^{n-1}]] = 0$ and $\Im[\tr[{\cal H}^{n-1}]] = 0$.

Based on these form of constraints, we can deduce that i) for $n = 2j, j \in \mathbb{Z}\backslash\{0\}$, aside from two constraints for $\det[H]=0$, $(n-2)$ constraints are for setting traces of even powers of $H$ to zero, i.e., $\tr[{\cal H}^{2l}]] = 0$ with $l<j$, and the remaining $(n-2)$ constraints enforce $\tr[{\cal H}^{2l+1}]] = 0$ with $l<j$. ii) For $n = 2j+1, j \in \mathbb{Z}\backslash\{0\}$, two constraints impose $\det[H]=0$, $(n-1)$ constraints ensures $\tr[{\cal H}^{2l}] = 0$ with $l<j$ and the rest of $(n-3)$ constraints impose $\tr[{\cal H}^{2l+1}] = 0$ with $l<j$.

In the following, we derive general statements for EP$n$s based on the energy constraints listed in Table~\ref{tab:symm} and using $\det[{\cal H}] = \prod_i \epsilon_i$ and $\tr[{\cal H}^k] = \sum_i \epsilon_i^k$ with $\epsilon$ be the eigenvalues of ${\cal H}$.

\subsection{Sublattice and pseudo-chiral symmetry}
In the presence of SLS or psCS, symmetry constraints enforce that $\{\epsilon({\bf k})\}=\{-\epsilon({\bf k})\}$. As a result, for $n=2j$ we get $\tr[{\cal H}^k]= \epsilon_1^k + \epsilon_2^k + \ldots + \epsilon_j^k + (-\epsilon_1)^k + (-\epsilon_2)^k + \ldots + (-\epsilon_j)^k = \epsilon_1^k + \epsilon_2^k + \ldots + \epsilon_j^k - \epsilon_1^k - \epsilon_2^k + \ldots -\epsilon_j^k = 0$, $\forall k \in \textrm{odd}$, while $\det[{\cal H}] \neq 0$ and $\tr[{\cal H}^k] \neq 0$, $\forall k \in \mathrm{even}$. Therefore, one needs to satisfy $2 + n - 2 = n$ constraints to find an EP$n$ with $n =2j$.

When $n = 2j+1$, at least one of the eigenvalues needs to be zero, such that $\det[{\cal H}] = 0$. We also find $\tr[{\cal H}^k]= 0$, $\forall k \in \mathrm{odd}$ as before. We thus are left with $n-1$ constraints that need to be satisfied to find an EP$n$ with $n = 2j+1$.

\subsection{Parity-time and pseudo-Hermitian symmetries}

In the presence of PT or psH symmetry, eigenvalues satisfy $\{\epsilon({\bf k})\}=\{\epsilon^*({\bf k})\}$. This implies that for $n = 2j+1$ at least one of the eigenvalues should be real. We save this real eigenvalue in $\epsilon_{j+1}$ for $n = 2j+1$ in the following.

For $n = 2j$, we find that $\det[{\cal H}] = \epsilon_1 \times \epsilon_2 \times \ldots \times \epsilon_j \times \epsilon_1^* \times \epsilon_2^* \times \ldots \times \epsilon_j^* = |\epsilon_1|^2 |\epsilon_2|^2 \ldots |\epsilon_j|^2 \in \mathbb{R}$, whereas for $n = 2j+1$, the determinant yields $\det[{\cal H}] = |\epsilon_1|^2 |\epsilon_2|^2 \ldots |\epsilon_j|^2 \epsilon_{j+1} \in \mathbb{R}$ with $\epsilon_{j+1} \in \mathbb{R}$. Similarly, using that $(c^*)^k = (c^k)^*$, we find for $n = 2j$ that $\tr[{\cal H}^k] = \epsilon_1^k + \epsilon_2^k + \ldots + \epsilon_j^k + (\epsilon_1^*)^k + (\epsilon_2^*)^k + \ldots + (\epsilon_j^*)^k = \epsilon_1^k + \epsilon_2^k + \ldots + \epsilon_j^k + (\epsilon_1^k)^* + (\epsilon_2^k)^* + \ldots + (\epsilon_j^k)^* = 2 \Re[\epsilon_1^k + \epsilon_2^k + \ldots + \epsilon_j^k] \in \mathbb{R}$ and for $n = 2j+1$ that $\tr[{\cal H}^k] = 2 \Re[\epsilon_1^k + \epsilon_2^k + \ldots + \epsilon_j^k] + \epsilon_{j+1} \in \mathbb{R}$. We thus conclude that $\Im[\det[{\cal H}]] = 0$ and $\Im[\tr[{\cal H}^k]] = 0$ generically, and we are left with $n-1$ constraints to realize EP$n$s, namely, $\Re[\det[{\cal H}]] = 0$ and $\Re[\tr[{\cal H}^k]] = 0$.

\subsection{Chiral and parity-particle-hole symmetries}
In the presence of CS or $\mathcal{CP}$ symmetry, eigenvalues display $\{\epsilon({\bf k})\}=\{-\epsilon^*({\bf k})\}$. We note that CS is not defined for $n = 2j+1$ as a result for odd dimensions in the following are merely relevant for the $\mathcal{CP}$ symmetry.
For $n = 2j+1$, we infer from the relation between the sets of eigenvalues that at least one of the eigenvalues is imaginary. We save this eigenvalue in $\epsilon_{j+1}$ for $n = 2j+1$.

For $n = 2j$, we find that $\det[{\cal H}] = \epsilon_1 \times \epsilon_2 \times \ldots \times \epsilon_j \times (-\epsilon_1^*) \times (-\epsilon_2^*) \times \ldots \times (-\epsilon_j^*) = (-1)^j |\epsilon_1|^2 |\epsilon_2|^2 \ldots |\epsilon_j|^2 \in \mathbb{R}$, whereas for $n = 2j+1$, we find that $\det[{\cal H}] = (-1)^j |\epsilon_1|^2 |\epsilon_2|^2 \ldots |\epsilon_j|^2 \epsilon_{j+1} \in \i \mathbb{R}$ with $\epsilon_{j+1} \in \i \mathbb{R}$.

For the traces we find for $n = 2j$ and $k\in \mathrm{odd}$ that $\tr[{\cal H}^k] = \epsilon_1^k + \epsilon_2^k + \ldots + \epsilon_j^k + (-\epsilon_1^*)^k + (-\epsilon_2^*)^k + \ldots + (-\epsilon_j^*)^k = \epsilon_1^k + \epsilon_2^k + \ldots + \epsilon_j^k + (-1)^k (\epsilon_1^k)^* + (-1)^k (\epsilon_2^k)^* + \ldots + (-1)^k (\epsilon_j^k)^* = \epsilon_1^k + \epsilon_2^k + \ldots + \epsilon_j^k - (\epsilon_1^k)^* - (\epsilon_2^k)^* + \ldots - (\epsilon_j^k)^* = 2 \i \Im[\epsilon_1^k + \epsilon_2^k + \ldots + \epsilon_j^k] \in \i \mathbb{R}$, while for $k \in \textrm{even}$, we get $\tr[{\cal H}^k] = \epsilon_1^k + \epsilon_2^k + \ldots + \epsilon_j^k + (-1)^k (\epsilon_1^k)^* + (-1)^k (\epsilon_2^k)^* + \ldots + (-1)^k (\epsilon_j^k)^* = \epsilon_1^k + \epsilon_2^k + \ldots + \epsilon_j^k + (\epsilon_1^k)^* + (\epsilon_2^k)^* + \ldots + (\epsilon_j^k)^* = 2 \Re[\epsilon_1^k + \epsilon_2^k + \ldots + \epsilon_j^k] \in \mathbb{R}$.

For $n=2j+1$, we find for $k\in \textrm{odd}$ that $\tr[{\cal H}^k] = 2 \i \Im[\epsilon_1^k + \epsilon_2^k + \ldots + \epsilon_j^k] + \epsilon_{j+1}^k \in i \mathbb{R}$, where we use that $\epsilon_{j+1}^k = \i^k (\Im[\epsilon_{j+1}])^k \in \i \mathbb{R}$ for odd $k$. For $k \in \textrm{even}$, we find $\tr[{\cal H}^k] = 2 \Re[\epsilon_1^k + \epsilon_2^k + \ldots + \epsilon_j^k] + \epsilon_{j+1}^k \in \mathbb{R}$, where we use that $\epsilon_{j+1}^k = \i^k (\Im[\epsilon_{j+1}])^k \in \mathbb{R}$ for even $k$.

For any $n = 2j$, we thus get $\det[{\cal H}] \in \mathbb{R}$, $\tr[{\cal H}^k] \in \mathbb{R}$ $\forall k\in \textrm{even}$ and $\tr[{\cal H}^k] \in \i  \mathbb{R}$ $\forall k\in \textrm{odd}$. This gives us $1 + (n-2)/2 + (n-2)/2 = n-1$ constraints. For $n = 2j+1$, we obtain $\det[{\cal H}] \in \i \mathbb{R}$, $\tr[{\cal H}^k] \in \mathbb{R}$, $\forall k\in \textrm{even}$ and $\tr[{\cal H}^k] \in \i  \mathbb{R}$, $\forall k\in \textrm{odd}$ leading to $1 + (n-1)/2 + (n-3)/2 = n-1$ constraints.

\section{Basis matrices for two-, three-, and four-band systems }

\subsection{Basis matrices for two-band systems} \label{app:pauli}

The basis matrices for two-band systems are Pauli matrices which read
\begin{align}
\sigma_{x} =
\begin{pmatrix}
0 & 1 \\
1 & 0
\end{pmatrix}
,\quad 
\sigma_{y} =
\begin{pmatrix}
0 & -\i \\
\i & 0
\end{pmatrix}
,\quad
\sigma_{z} =
\begin{pmatrix}
1 & 0 \\
0 & -1
\end{pmatrix}.
\end{align}

\subsection{Basis matrices for three-band systems} \label{app:gellmann_su3}

The basis matrices for three-band systems are the Gell-Mann matrices, that span the Lie algebra of the SU(3) group,
\begin{align}
\tr[M^{\alpha}]= 0 \quad \forall \alpha \in \{ 1,\ldots 8\}, \, \text{with }
(M^{\alpha})^{\dagger} = M^{\alpha}, 
\end{align}
\begin{align}
M^{1}&=
 \begin{pmatrix}
0  & -\i & 0 \\
\i & 0 & 0 \\
0 & 0 & 0
\end{pmatrix}, \quad 
M^{2}=
 \begin{pmatrix}
0  & 0& -\i  \\
0& 0 & 0 \\
\i & 0 & 0
\end{pmatrix} ,\\ 
M^{3}&= 
\begin{pmatrix}
0  & 0 & 0 \\
0 & 0 & -\i \\
0 & \i & 0
\end{pmatrix}, \quad 
M^{4}= 
\begin{pmatrix}
0  & 1 & 0 \\
1 & 0 & 0 \\
0 & 0 & 0
\end{pmatrix}, \\
M^{5}&= 
 \begin{pmatrix}
0  & 0 & 1 \\
0 & 0 & 0 \\
1 & 0 & 0
\end{pmatrix}, \quad 
M^{6}=
 \begin{pmatrix}
0  & 0 & 0 \\
0 & 0 & 1 \\
0 & 1 & 0
\end{pmatrix},\\
M^{7}&=
  \begin{pmatrix}
1 & 0 & 0 \\
0 & -1 & 0  \\
0 & 0 & 0
\end{pmatrix}, \quad 
M^{8}= 
\begin{pmatrix}
\frac{1}{\sqrt{3}}  & 0 & 0 \\
0 & \frac{1}{\sqrt{3}} & 0 \\
0 & 0 & -\frac{2}{\sqrt{3}}
\end{pmatrix}.
\end{align}
These matrices satisfy (anti-)commutation relations and the $SU(3)$ Fierz completeness relations
\begin{align}
[M^{\alpha}, M^{\beta}] &= 2 \i f_{\alpha \beta \gamma} M^{\gamma} ,\label{eq:commuteM}
\\
\{ M^{\alpha}, M^{\beta}\} &=\frac{4}{3} \delta_{\alpha \beta} \id
+2 d_{\alpha \beta \gamma} M^{\gamma}, 
\label{eq:anticommuteM}
\\
\delta_{i l} \delta_{kj}
&=
\frac{1}{3} \delta_{ij} \delta_{kl}
+
\frac{1}{2} M^{\alpha}_{ij} M^{\alpha}_{kl},
\label{eq:Fcr} 
\\
M^{\alpha}_{ij} M^{\alpha}_{kl}  &=
 \frac{16}{9} \delta_{il} \delta_{kj} - \frac{1}{3} M^{\alpha}_{il} M^{\alpha}_{kj}.\label{eq:MM_to_MM}
\end{align}
Here $d_{abc}$~($f_{abc}$) are the (anti-)symmetric structure constant of SU(3)~\cite{Georgi1982,Barnett2012}.

\subsection{Basis matrices for four-band systems} \label{app:gellmann_su4}

The basis matrices for four-band systems are the generalized Gell-Mann matrices, that span the Lie algebra of the SU(4) group,
\begin{align}
\tr[\Lambda^{\alpha}]= 0 \quad \forall \alpha \in \{ 1,\ldots 15\}, \, \text{with }
(\Lambda^{\alpha})^{\dagger} = \Lambda^{\alpha}, 
\end{align}
\begin{align}
\Lambda^{1}&= 
\begin{pmatrix}
0 & -\i & 0 & 0 \\
\i & 0 & 0 & 0 \\
0 & 0 & 0 & 0 \\
0 & 0 & 0 & 0 \\
\end{pmatrix}
,\quad 
\Lambda^{2}= 
\begin{pmatrix}
0 & 0 &-\i  & 0 \\
0 & 0 & 0 & 0 \\
\i & 0 & 0 & 0 \\
0 & 0 & 0 & 0 \\
\end{pmatrix}
,\\
\Lambda^{3}&= 
\begin{pmatrix}
0 & 0 & 0 &-\i   \\
0 & 0 & 0 & 0 \\
0 & 0 & 0 & 0 \\
\i & 0 & 0 & 0 \\
\end{pmatrix}
,\quad 
\Lambda^{4}= 
\begin{pmatrix}
0 & 0 & 0 &0   \\
0 & 0 & -\i & 0 \\
0 & \i & 0 & 0 \\
0 & 0 & 0 & 0 \\
\end{pmatrix}
,\\
\Lambda^{5}&= 
\begin{pmatrix}
0 & 0 & 0 &0   \\
0 & 0 & 0 & -\i \\
0 & 0 & 0 & 0 \\
0 & \i & 0 & 0 \\
\end{pmatrix}
,\quad 
\Lambda^{6}= 
\begin{pmatrix}
0 & 0 & 0 &0   \\
0 & 0 & 0 & 0 \\
0 & 0 & 0 & -\i \\
0 & 0 & \i & 0 \\
\end{pmatrix}
,\\ 
\Lambda^{7}&= 
\begin{pmatrix}
0 & 1 & 0 &0   \\
1 & 0 & 0 & 0 \\
0 & 0 & 0 & 0 \\
0 & 0 & 0 & 0 \\
\end{pmatrix}
,\quad 
\Lambda^{8}= 
\begin{pmatrix}
0 & 0 & 1 &0   \\
0 & 0 & 0 & 0 \\
1 & 0 & 0 & 0 \\
0 & 0 & 0 & 0 \\
\end{pmatrix}
,\\
\Lambda^{9}&= 
\begin{pmatrix}
0 & 0 & 0 &1   \\
0 & 0 & 0 & 0 \\
0 & 0 & 0 & 0 \\
1 & 0 & 0 & 0 \\
\end{pmatrix}
,\quad
\Lambda^{10}= 
\begin{pmatrix}
0 & 0 & 0 & 0   \\
0 & 0 & 1 & 0 \\
0 & 1 & 0 & 0 \\
0 & 0 & 0 & 0 \\
\end{pmatrix}
,\\
\Lambda^{11}&= 
\begin{pmatrix}
0 & 0 & 0 & 0   \\
0 & 0 & 0 & 1 \\
0 & 0 & 0 & 0 \\
0 & 1 & 0 & 0 \\
\end{pmatrix}
,\quad
\Lambda^{12}= 
\begin{pmatrix}
0 & 0 & 0 & 0   \\
0 & 0 & 0 & 0 \\
0 & 0 & 0 & 1 \\
0 & 0 & 1 & 0 \\
\end{pmatrix}
,\\
\Lambda^{13}&= 
\begin{pmatrix}
1 & 0 & 0 & 0   \\
0 & -1 & 0 & 0 \\
0 & 0 & 0 & 0 \\
0 & 0 & 0 & 0 \\
\end{pmatrix}
,\,
\Lambda^{14}= 
\begin{pmatrix}
\frac{1}{\sqrt{3}} & 0 & 0 & 0   \\
0 & \frac{1}{\sqrt{3}} & 0 & 0 \\
0 & 0 & -\frac{2}{\sqrt{3}} & 0 \\
0 & 0 & 0 & 0 \\
\end{pmatrix}
,\\
\Lambda^{15}&= 
\begin{pmatrix}
\frac{1}{\sqrt{6}} & 0 & 0 & 0   \\
0 & \frac{1}{\sqrt{6}} & 0 & 0 \\
0 & 0 & \frac{1}{\sqrt{6}} & 0 \\
0 & 0 & 0 & -\sqrt{\frac{3}{2}} \\
\end{pmatrix}.
\end{align}
Aside from the above matrices, one can use the $\Gamma$ matrices, basis of the $\Gamma-$group, as
\begin{align}
\Gamma_{1} = \sigma_{x} \otimes \tau_{0} = \Lambda^{8} + \Lambda^{11}=
\left(
\begin{array}{cccc}
 0 & 0 & 1 & 0 \\
 0 & 0 & 0 & 1 \\
 1 & 0 & 0 & 0 \\
 0 & 1 & 0 & 0 \\
\end{array}
\right),
\\
\Gamma_{2} = \sigma_{y} \otimes \tau_{y} = \Lambda^{10} - \Lambda^{9}=
\left(
\begin{array}{cccc}
 0 & 0 & 0 & -1 \\
 0 & 0 & 1 & 0 \\
 0 & 1 & 0 & 0 \\
 -1 & 0 & 0 & 0 \\
\end{array}
\right)
,\\
\Gamma_{3} = \sigma_{z} \otimes \tau_{0} = \frac{2}{\sqrt{3}} \Lambda^{14} + \sqrt{\frac{2}{3} } \Lambda^{15}
=
\left(
\begin{array}{cccc}
 1 & 0 & 0 & 0 \\
 0 & 1 & 0 & 0 \\
 0 & 0 & -1 & 0 \\
 0 & 0 & 0 & -1 \\
\end{array}
\right)
,\\
\Gamma_{4} = \sigma_{y} \otimes \tau_{x} = \Lambda^{3} + \Lambda^{4}
=
\left(
\begin{array}{cccc}
 0 & 0 & 0 & -\i \\
 0 & 0 & -\i & 0 \\
 0 & \i & 0 & 0 \\
 \i & 0 & 0 & 0 \\
\end{array}
\right)
,\\
\Gamma_{5} = \sigma_{y} \otimes \tau_{z} = \Lambda^{2} - \Lambda^{5}=
\left(
\begin{array}{cccc}
 0 & 0 & -\i & 0 \\
 0 & 0 & 0 & \i \\
 \i & 0 & 0 & 0 \\
 0 & -\i & 0 & 0 \\
\end{array}
\right).
\end{align}
The above $\Gamma$ matrices satisfy the Clifford algebra such that $\{ \Gamma^{
\mu}, \Gamma^{\nu} \} = 2 \eta^{\mu \nu} \id_{4 \times 4}$ with $\mu , \nu \in \{1, \ldots,4\}$. $\eta^{
\mu \nu}$ denotes the metric signature of the space, i.e., Minkowski or Euclidean signatures. Using $\Gamma-$matrices in Hermitian systems implies that we have spatial rotations and Lorentz boosts in these systems.

\section{Generic eigenvalue solutions for a three-band model} \label{app:eigs_three_band}

Here we present the solutions to the fourth-order characteristic polynomial in Eq.~\eqref{eq:char_poly_3}, which are the eigenvalues to the Hamiltonian in Eq.~\eqref{eq:H4d}. These solutions read
\begin{widetext}
\begin{align}
\lambda_{1} &=
\frac{1}{6} \left(2^{2/3} \sqrt[3]{\sqrt{4 \eta ^3+\nu ^2}+\nu }-\frac{2 \sqrt[3]{2} \eta }{\sqrt[3]{\sqrt{4 \eta ^3+\nu ^2}+\nu }}+2 \tr[{\cal H}]\right)
\label{eq:lam1-lvl3}
,\\
\lambda_{2} &=
\frac{1}{12} \left(2^{2/3} \i \left(\i +\sqrt{3}\right) \sqrt[3]{\sqrt{4 \eta ^3+\nu ^2}+\nu }+\frac{\sqrt[3]{2} \left(2+2 \i \sqrt{3}\right) \eta }{\sqrt[3]{\sqrt{4 \eta ^3+\nu ^2}+\nu }}+4 \tr[{\cal H}]\right)
\label{eq:lam2-lvl3}
,\\
\lambda_{3} &=
\frac{1}{12} \left(2^{2/3} \i \left(\i- \sqrt{3}\right) \sqrt[3]{\sqrt{4 \eta ^3+\nu ^2}+\nu }+\frac{\sqrt[3]{2} \left(2-2 \i \sqrt{3}\right) \eta }{\sqrt[3]{\sqrt{4 \eta ^3+\nu ^2}+\nu }}+4 \tr[{\cal H}]\right)
\label{eq:lam3-lvl3}
.
\end{align}
\end{widetext}
Here $\eta$ and $\nu$ are defined in Eqs.~(\ref{eq:eta3lvl}, \ref{eq:nu3lvl}) in the main text.

\section{Generic eigenvalue solutions for a four-band model} \label{app:eigs_four_band}

Here we present the solutions to the fourth-order characteristic polynomial in Eq.~\eqref{eq:char_poly_4}, which are the eigenvalues to the Hamiltonian in Eq.~\eqref{eq:H4d}. These solutions read
\begin{widetext}
\begin{align}
    \lambda_{1} =&
\frac{ -\sqrt{6} \sqrt{-\frac{3 \sqrt{3} \kappa }{\sqrt{3 a^2-8 b+2\ 2^{2/3} \sqrt[3]{\sqrt{\nu ^2-4 \eta ^3}+\nu }+\frac{4 \sqrt[3]{2} \eta }{\sqrt[3]{\sqrt{\nu ^2-4 \eta ^3}+\nu }}}}+3 a^2-8 b-2^{2/3} \sqrt[3]{\sqrt{\nu ^2-4 \eta ^3}+\nu }-\frac{2 \sqrt[3]{2} \eta }{\sqrt[3]{\sqrt{\nu ^2-4 \eta ^3}+\nu }}}}{12}
\nonumber \\
&
-
\frac{
\sqrt{3} \sqrt{3 a^2-8 b+2^{5/3} \sqrt[3]{\sqrt{\nu ^2-4 \eta ^3}+\nu }+\frac{4 \sqrt[3]{2} \eta }{\sqrt[3]{\sqrt{\nu ^2-4 \eta ^3}+\nu }}}+3 a }{12}
\label{eq:lam1-lvl4}
,\\
\lambda_{2} &=
\frac{ \sqrt{6} \sqrt{-\frac{3 \sqrt{3} \kappa }{\sqrt{3 a^2-8 b+2\ 2^{2/3} \sqrt[3]{\sqrt{\nu ^2-4 \eta ^3}+\nu }+\frac{4 \sqrt[3]{2} \eta }{\sqrt[3]{\sqrt{\nu ^2-4 \eta ^3}+\nu }}}}+3 a^2-8 b-2^{2/3} \sqrt[3]{\sqrt{\nu ^2-4 \eta ^3}+\nu }-\frac{2 \sqrt[3]{2} \eta }{\sqrt[3]{\sqrt{\nu ^2-4 \eta ^3}+\nu }}}
}{12}
\nonumber \\
&
-
\frac{\sqrt{3} \sqrt{3 a^2-8 b+2^{5/3} \sqrt[3]{\sqrt{\nu ^2-4 \eta ^3}+\nu }+\frac{4 \sqrt[3]{2} \eta }{\sqrt[3]{\sqrt{\nu ^2-4 \eta ^3}+\nu }}}+3 a }{12}
\label{eq:lam2-lvl4}
,\\
\lambda_{3} &=
\frac{ -\sqrt{6} \sqrt{\frac{3 \sqrt{3} \kappa }{\sqrt{3 a^2-8 b+2\ 2^{2/3} \sqrt[3]{\sqrt{\nu ^2-4 \eta ^3}+\nu }+\frac{4 \sqrt[3]{2} \eta }{\sqrt[3]{\sqrt{\nu ^2-4 \eta ^3}+\nu }}}}+3 a^2-8 b-2^{2/3} \sqrt[3]{\sqrt{\nu ^2-4 \eta ^3}+\nu }-\frac{2 \sqrt[3]{2} \eta }{\sqrt[3]{\sqrt{\nu ^2-4 \eta ^3}+\nu }}}
}{12}
\nonumber \\
&
+
\frac{\sqrt{3} \sqrt{3 a^2-8 b+2^{5/3} \sqrt[3]{\sqrt{\nu ^2-4 \eta ^3}+\nu }+\frac{4 \sqrt[3]{2} \eta }{\sqrt[3]{\sqrt{\nu ^2-4 \eta ^3}+\nu }}}+3 a }{12}
\label{eq:lam3-lvl4}
,\\
\lambda_{4} &=
\frac{ \sqrt{6} \sqrt{\frac{3 \sqrt{3} \kappa }{\sqrt{3 a^2-8 b+2\ 2^{2/3} \sqrt[3]{\sqrt{\nu ^2-4 \eta ^3}+\nu }+\frac{4 \sqrt[3]{2} \eta }{\sqrt[3]{\sqrt{\nu ^2-4 \eta ^3}+\nu }}}}+3 a^2-8 b-2^{2/3} \sqrt[3]{\sqrt{\nu ^2-4 \eta ^3}+\nu }-\frac{2 \sqrt[3]{2} \eta }{\sqrt[3]{\sqrt{\nu ^2-4 \eta ^3}+\nu }}}
}{12}
\nonumber \\
&
+
\frac{\sqrt{3} \sqrt{3 a^2-8 b+2^{5/3} \sqrt[3]{\sqrt{\nu ^2-4 \eta ^3}+\nu }+\frac{4 \sqrt[3]{2} \eta }{\sqrt[3]{\sqrt{\nu ^2-4 \eta ^3}+\nu }}}+3 a }{12}
\label{eq:lam4-lvl4}
.
\end{align}
\end{widetext}
Here $\eta$, $\nu$ and $\kappa$ are defined in Eqs.~(\ref{eq:eta-4band}, \ref{eq:nu-4band}, \ref{eq:kappa-4band}) in the main text.

\end{document}